\documentclass[12pt,preprint]{aastex}

\newcommand\spitzer{\textit{Spitzer}}
\newcommand\inv{$^{-1}$}
\newcommand{\ltsimeq}{\raisebox{-0.6ex}{$\,\stackrel
        {\raisebox{-.2ex}{$\textstyle <$}}{\sim}\,$}}
\newcommand{\gtsimeq}{\raisebox{-0.6ex}{$\,\stackrel
        {\raisebox{-.2ex}{$\textstyle >$}}{\sim}\,$}}

\slugcomment{Astrophysical J. (in press Dec2007)}

\received{11 May 2007}
\revised{01 August 2007}
\accepted{}

\shorttitle{Water in C/2003 K4 (LINEAR)}
\shortauthors{Woodward et al.}

\begin{document}



\title{WATER IN COMET C/2003~K4 (LINEAR) WITH {\it SPITZER}}


\author{
CHARLES E. WOODWARD\altaffilmark{1},
MICHAEL S. KELLEY\altaffilmark{1,2}, \\
DOMINIQUE BOCKEL\'EE-MORVAN\altaffilmark{3},\\
R.~D. GEHRZ\altaffilmark{1}
}


\altaffiltext{1}{Department of Astronomy, School of Physics and 
Astronomy, 116 Church Street, S.~E., University of Minnesota, 
Minneapolis, MN 55455,\ \it{chelsea@astro.umn.edu, gehrz@astro.umn.edu}} 

\altaffiltext{2}{Current address: Department of Physics, University
of Central Florida, 4000 Central Florida Blvd., Orlando, FL
32816-2385, \ \it{msk@physics.ucf.edu} }

\altaffiltext{3}{LESIA, Observatoire de Paris, 5 place Jules Janssen, 
F92195, Meudon, France, \\ \it{dominique.bockelee@obspm.fr}}


\begin{abstract}

We present sensitive 5.5--7.6~\micron\mbox{} spectra of comet 
C/2003~K4 (LINEAR) obtained on 16 July 2004 ($r_{h} = 1.760$~AU,
$\Delta_{Spitzer} = 1.409$~AU, phase angle 35.4\degr) with the 
\spitzer~Space Telescope. The $\nu_{2}$ vibrational band of water is
detected with a high signal-to-noise ratio (\gtsimeq 50). Model 
fitting to the best spectrum yields a water ortho-to-para 
ratio of $2.47\pm0.27$, which corresponds to a spin 
temperature of $28.5^{+6.5}_{-3.5}$~K. Spectra 
acquired at different offset positions show that the rotational
temperature decreases with increasing distance from the nucleus, which is 
consistent with evolution from thermal to fluorescence equilibrium.
The inferred water production rate is $(2.43\pm0.25) \times 
10^{29}$\, molec.~s$^{-1}$. The spectra do not show any evidence 
for emission from PAHs and carbonate minerals, in contrast to results 
reported for comets 9P/Tempel 1 and C/1995 O1 (Hale-Bopp). However, 
residual emission is observed near 7.3~\micron\mbox{} the
origin of which remains unidentified.   
 
\end{abstract}



\section{INTRODUCTION\label{intro}}

The composition of cometary nuclei probes the physical conditions in the 
early solar nebula, the survival of materials from the interstellar medium 
(ISM), and the cold dense molecular cloud core in which the solar system 
formed \citep{wooden04, ehren04, mum03}. Comet nuclei are highly porous 
agglomerates of ice and dust grains, perhaps with highly stratified, 
inhomogeneous layers of varied density, porosity, and composition 
\citep{harker07, belton06, oro06, ahearn05, prialnik04}. The nucleus 
composition is dominated by ices (primarily water), 
organic refractory materials, silicates, and carbonaceous materials. When 
comets are within heliocentric distances of $r_{h} \leq 20$~AU,
solar insolation triggers sublimation and the release of volatile 
gases, sometimes sporadically, forming observable comae \citep{meechs04}. 

In the nucleus of a comet, volatiles are frozen as ices or trapped as 
gases in amorphous water ice \citep{capria02, prialnik02}. Cometary 
activity occurs when gases are released through sublimation or through 
the exoergic crystallization of amorphous water ice. Between $\sim 20$
to 5~AU, when nuclear surface temperatures reach $\simeq 20 - 100$~K, 
CO ice sublimes from 
the nucleus \citep{capria00, prialnik02}, possibly from near the surface 
\citep{gunna03}, and triggers activity and intermittent 
outbursts. Between $\sim 6$ to $\sim 4$~AU, a dramatic increase occurs in 
gas production and grain entrainment and signals the coma onset stage. At 
nuclear surface temperatures of $\sim 120 - 130$~K \citep{ prialnik04}, the 
water ice phase transition (amorphous to crystalline) releases a 
fraction of the trapped volatile gases. Strong erosion maintains 
the CO-ice sublimation and phase transition 
fronts relatively close to the surface \citep{capria00}. At $\sim 4$ to 
$\sim 3$~AU, the near-surface crystalline water ice layer, with its 
remaining trapped gases, begins to sublime. Water sublimation drives this 
vigorous activity stage that is often characterized by discrete active 
areas or jets. 

Water is the dominate ice in comet nuclei and the production rate of
water is correlated with comet activity. It influences the 
thermal balance of the coma as a strong coolant.  At some wavelengths $\leq
10$~\micron \ emission from ro-vibrational transitions of water can 
dominate the spectral energy distribution \citep{cro97b}; water can also
be observed from its rotational transitions at submillimeter wavelengths 
\citep[see review of][]{boc04}. Probed through spectroscopic observations of
coma species, the water production rate, coma temperature, and the 
nuclear spin temperatures derived from ortho-to-para ratios (OPR) 
are of particular interest in the study of cometary atmospheres and 
cometary physics. These physical characteristics, complemented 
by knowledge of the nucleus refractory and ice composition, 
provide constraints on solar nebula models
\citep{mum03,marcharn04}, and restrict the formation zones within the
protoplanetary disk where cometary nuclei could conglomerate.  In
particular, the nuclear spin temperature of water measured in 
comet comae may be indicative of the chemical formation 
temperature of water \citep{del05, mum93} therefore identifying 
the environment where pre-cometary ices condensed.

Here we present longslit {\it Spitzer Space Telescope} \ spectroscopic
observations of the 6~\micron \ $\nu_{2}$ vibrational band of water detected
in comet C/2003 K4 (LINEAR) at $r_{h} = 1.760$~AU. The high
signal-to-noise and the Infrared Spectrograph (IRS) longslit
enable us to extract spatially resolved spectra in the coma and to 
measure the water production rate, $Q$(H$_2$O), and  
the rotational temperature, $T_{rot}$, and the OPR variation in the
coma. Space observations of the strong $\nu_{2}$ fundamental 
bands near 6~\micron \ present a potentially more advantageous 
method for constraining water production rates and $T_{rot}$ in 
comets than the more common ground-based measurement of the weaker 
non-resonance fluorescent ``hot-bands'' near 2.9~\micron \ as
the complex corrections for telluric extinction, slit-loss due to 
seeing, and consideration of whether the local radiative pump in the coma 
is optically thick are minimized \citep{bonev07, bonev06, del04, dbm87}. 
In addition, accurate laboratory 
measurements of the absorption line strengths used to compute Einstein
coefficients, $A_{\nu^{\prime},\nu^{\prime\prime}}$(s$^{-1}$), 
for the $\nu_{2}$ pump from the ground-state (000) are 
extant \citep{BT2, del04, PS97}
while those for the hot bands are more challenging, leading to
some uncertainty in estimates of the spontaneous emission rates,
$g_{\nu^{\prime},\nu^{\prime\prime}}$~(s$^{-1}$).

The infrared $\nu_{2}$ band of water was first detected with the
Short Wavelength Spectrometer (SWS) of the \textit{Infrared Space
Observatory} (ISO) in the exceptional comet C/1995 O1 (Hale-Bopp) at
$r_{h} \simeq 2.9$~AU \citep{cro97a}.  The SWS spectral resolution of
$\sim$ 1000 resulted in the detection of several individual
ro-vibrational lines.  However, the low signal-to-noise ratio
prevented detailed analysis of their relative intensities.  We discuss
the \spitzer{} observations and data reduction techniques in
\S~\ref{obs}.  Section~\ref{disc-mfit} discusses the modeling of the
$\nu_{2}$ water band. Section~\ref{disc-h20} through
\S~\ref{h20rate} present the results, followed in  
\S~\ref{residuals} by a discussion of residual emission features, including 
a comparison with the $\nu_{2}$ band of water detected in other comets 
with \spitzer{}. Section~\ref{concl} presents a summary of our study of 
comet C/2004~K4 (LINEAR).

\section{OBSERVATIONS AND REDUCTION\label{obs}}

First identified as an asteroidal object in the LINEAR survey, 
C/2003 K4 (LINEAR) was discovered to have an extended spherical coma 
by \citet{iauc8131} with parabolic orbital 
elements consistent with that of a dynamically new Oort Cloud comet. 
The following year upon perihelion approach ($q = 1.02$~AU, 
2004 Oct 13.5~UT), C/2003 K4 (LINEAR) was bright in the optical 
(V \ltsimeq 7~mag) and was noted to exhibit a primarily featureless 
10~\micron \ spectral energy distribution with emission
from large amorphous carbon and silicate grains (grain radii 
$\ge 0.7$~\micron) dominating the coma \citep{iauc8378}. The 
10~\micron \ silicate feature-to-continuum ratio was observed to be 
near unity \citep{iauc8391, iauc8361}, with little evidence for structure 
near 11.2~\micron \ attributable to Mg-rich crystalline olivine 
grains \citep{iauc8391, iauc8378}. \citet{shulz05} observed a 
single broad coma feature in broadband images, perpendicular to the sunward 
direction in C/2003 K4 (LINEAR) during the 2004 May through 2004 
July period with an increase 
in the amount of inferred gas-contamination in the B and V coma 
surface brightness with decreasing heliocentric distance. During the 
epoch of our \spitzer{} observations (\S~\ref{obs-irs}), the comet produced 
a considerable amount of dust \citep[$Af\rho \simeq 10,000$~cm;][]{shulz05}.

\subsection{Spitzer IRS\label{obs-irs}}

Spectra of comet C/2003 K4 (LINEAR) were obtained with the Infrared
Spectrograph (IRS) instrument \citep{houck04} on the \textit{Spitzer
Space Telescope} \citep{gehrz07, werner04}.  The comet was observed in
the second order of the short-wavelength, low-resolution module (SL2)
on 2004 Jul 16 at 04:56~UT as part of a \spitzer{} Guaranteed Time 
Observation (GTO) program (PI: R.D. Gehrz), program 
identification (PID) 131, astronomical observation request (AOR) key
\dataset[ADS/Sa.Spitzer#0008525056]{0008525056}, and processed with
IRS reduction pipeline S15.3.0.  The SL2 slit is 3.7\arcsec{} wide
and provides 57\arcsec{} of spatially resolved spectra
(1.8\arcsec~pixel$^{-1}$) with a spectral dispersion of 0.06~\micron.
Six spectra (14~s $\times$ 3 cycles) at 5.2--7.6~\micron{} were
recorded in a $3\times2$ spectral map, with
$7.2\arcsec\times78\arcsec$ \ steps (perpendicular $\times$ parallel to
the long slit dimension).  The comet was at a heliocentric distance
($r_h$) of 1.760~AU, a \spitzer{}-comet distance of 1.409~AU, and a
phase angle of 35.4\degr.

\begin{figure}
\plotone{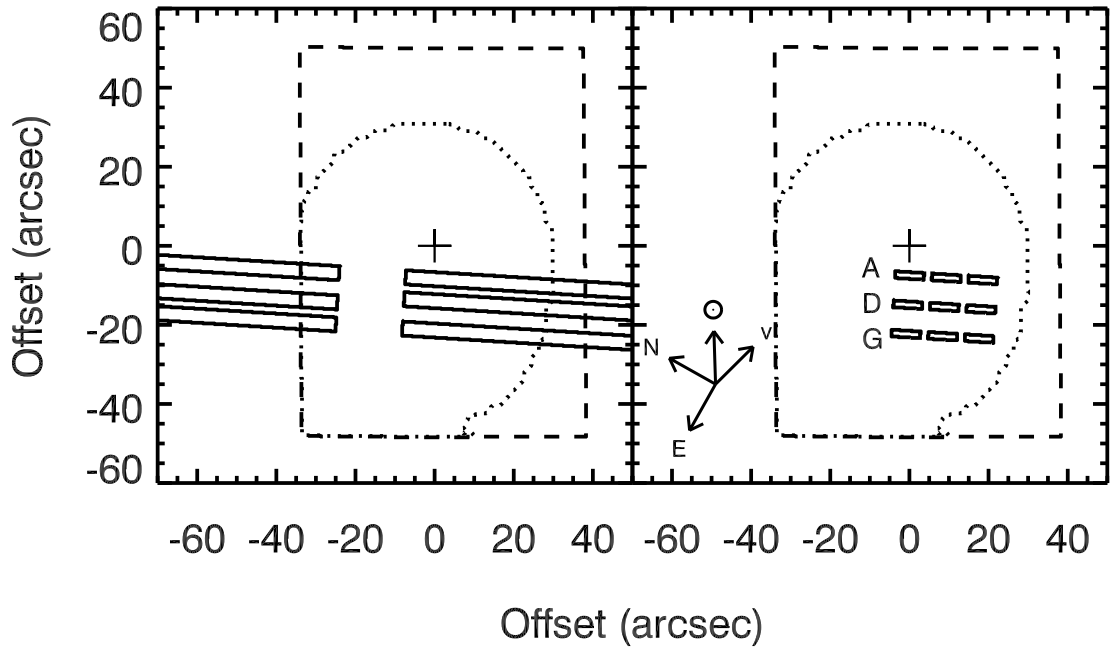}
\caption{\textbf{Left:} The 15~\micron{} IRS acquisition image and 
observed slit positions in comet C/2003 K4 (LINEAR).  The 
\textit{dotted contour} outlines the 
saturation ellipse in the image and the \textit{dashed line} traces the 
edge of the peak-up array.  The JPL \#96 ephemeris position of the 
comet nucleus is marked with a \textit{cross} (see 
\S~\ref{obs}).  The slit positions are outlined with \textit{solid 
lines}. Celestial north (N) and east (E), the comet's heliocentric 
velocity (v), and the direction of the sun ($\sun$) are marked with 
\textit{arrows}.  \textbf{Right:} The nine slit extraction apertures are 
outlined with \textit{solid lines} and run from left to right, and top to 
bottom (see Table~\ref{table:tb_k4}) A, B, C (row 1), D, E, F 
(row 2), and G, H, I (row 3). 
\label{fig:slit}} 
\end{figure}

At the time of acquisition, we attempted to acquire the comet nucleus
with the \spitzer{} IRS 15~\micron{} peak-up array.  However, the
bright inner coma saturated a $76\arcsec\times64\arcsec$ ellipse in
the $98\arcsec\times72\arcsec$ peak-up array preventing the spacecraft
from computing a centroid on the comet. Thus, the telescope pointed to
the comet's nominal ephemeris position derived from orbital elements
uploaded to the spacecraft prior to the execution of the AOR. On the
date of observation, 2004 Jul 16~UT, the nominal position of the comet
derived from these elements was 28\arcsec{} from the actual position
calculated with revised elements from JPL ephemeris \#96 (computed
2006 Dec 14 with a data-arc spanning from 2003 May 28 through 2006 Nov
17). However, comet C/2003 K4 (LINEAR) had an extensive 
coma ($\ge 1$\arcmin \ in diameter) at the epoch of our 
\spitzer{} observation and thus error
in the position of the nucleus did not affect our ability to obtain
spectra of the comet coma. Our discussion of the IRS slit positions within
the coma of the comet are referenced to the actual position as
computed from the most recent JPL ephemeris. Figure~\ref{fig:slit}
shows the blue peak-up image (saturated core), the slit positions, and
the position of the nucleus (\textit{cross}).

Coma emission (including the spectral signature from water lines)
is present in all portions of our slits to varying
degrees. Therefore, a robust estimate of the background emission is
difficult to accurately assess (to the level of a few percent) using
the \spitzer{} longslit observations of comet 
C/2003 K4 (LINEAR) alone.  However, in the same IRS campaign 
(\#10) an observation with the same
IRS AOR parameters toward a similar ecliptic latitude 
($52.5\degr \pm 0.5\degr$) was available. The background 
derived from this IRS observation
(AOR key \dataset[ADS/Sa.Spitzer#0004733952]{0004733952} 
obtained from the \spitzer{} archive) was two-dimensionally 
subtracted from the basic calibrated data products (BCDs)
 of comet C/2003 K4 (LINEAR).

After background subtraction, we corrected the world coordinate system
of the two-dimensional spectral frames for the motion of 
the comet, then combined
each source observation into data cubes with the 
CUbe Builder for IRS Spectra Maps \citep[CUBISM,][]{smith07}  
program (v1.5)\footnote{Available at
http://ssc.spitzer.caltech.edu/archanaly/contributed/cubism/}. 
CUBISM combines each cycle and each 
slit position into a data cube where two
axes contain the spatial information (1.85\arcsec~pixel\inv{} grid),
and the remaining axis contains the spectral information.  A separate
cube is created for the pipeline errors derived from 
the individual BCDs.  The program photometrically 
calibrates the data, including a correction for diffraction 
losses at the entrance slit
\citep[the so-called slit-loss correction factor;][]{kelley06, irsdh}.

IRS spectra extracted from SL2 module have weak fringing 
artifacts (\ltsimeq 2\% of the source flux, $F(\lambda_{i})$, at a given 
wavelength) that are difficult to completely remove 
using a sinusoidal function, as 
they are not spectrally resolved and vary with position in the 
slit \citep{irsdh}. Thus, to account for any potential 
residual fringe contamination we 
compute the flux uncertainty in our extracted spectra at a given 
wavelength from the quadrature sum of the photometric error arising from 
the pipeline processing of individual BCDs plus a contribution due to SL2 
fringing signal equal to $0.02 \times F(\lambda_{i})$. This latter term
is an {\it upper limit} to the fringe uncertainty.

We extracted spectra from nine locations in the coma, as shown in
Fig.~\ref{fig:slit}. The extraction apertures are
$1.85\arcsec \times 7.40\arcsec$ rectangles (subtending 1890~km
$\times$ 7560~km within the coma). We restricted our nine source
extractions and subsequent analysis to locations where the coma is 
brightest, from 0\arcsec \ to $+30$\arcsec \ offset from the nucleus 
(Fig.~\ref{fig:slit}).

\subsection{Wavelength Calibration\label{obs-wave}}

The water lines are unresolved and blended in the IRS
spectra.  Furthermore, the \spitzer{} IRS spectra are calibrated with
standards measured near the center of the slit and three of our
extractions occur near the slit edge. 
Analysis of the wavelength calibration and unresolved line widths is
critical for identifying and fitting the water lines in the IRS
spectra. To verify the wavelength calibration at the slit edge,
we reduced IRS calibration observations of NGC~7027 at the center
and edge of the SL2 slit.  The NGC~7027 observations (AOR key
\dataset[ADS/Sa.Spitzer#0010066432]{0010066432}, \spitzer{} PID
1410, IRS pipeline S13.2.0) were taken during the same IRS observing
campaign (IRS\#10) as the C/2003 K4 (LINEAR) 
spectra.  We fit Gaussians to the
[\ion{Mg}{5}] 5.61~\micron{} line \citep{bernard01} with the nebula at
the center and edge positions.  The width of the [\ion{Mg}{5}]
line in NGC 7027 at the center position agrees with the spectral
resolution solution provided by the \spitzer{} Science Center
(0.0605~\micron).  At the edge position, the line width increases to
$0.0655\pm0.0002$~\micron{}. The [\ion{Mg}{5}] line was observed
at a central wavelength of $5.6242\pm0.0003$~\micron{}.  Accounting
for Doppler shift, the observed wavelength is
$0.0137\pm0.0007$~\micron \ from the vacuum rest wavelength of
$5.6099\pm0.0006$~\micron.  The IRS SL2 wavelength calibration is
$\pm0.006$~\micron{} (r.m.s.), indicating the [\ion{Mg}{5}] shift is
2.3-$\sigma$ from the expected central wavelength. The Gaussian
fits show no significant difference in central wavelength between the
center and edge positions. We compare the wavelength positions of
the water lines to the expected central wavelengths in \S~\ref{dbm_anl}.
 
\section{THE $\nu_2$ WATER BAND IN COMET C/2003 K4 (LINEAR)\label{dbm_anl}}

\subsection{Model Fitting\label{disc-mfit}}

At the resolution of the SL2 \spitzer{} spectrometer ($R \approx
100$), the $\nu_{2}$ water 
band shows ro-vibrational structure from which information on the 
rotational temperature, $T_{rot}$, in the ground vibrational state can be 
obtained. Though the spectral resolution is not high enough to separate 
individual ro-vibrational lines, and therefore ortho from para 
water lines, it is still possible to assess whether our \spitzer{} 
spectra can provide some constraints on the OPR. Previous 
determinations of the OPR in cometary 
water were based on water infrared 
spectra obtained with resolving powers between 1500 and 25000 
\citep[e.g.,][]{mum93,cro97b,del05,kawakita06,bonev07}. 

Vibrational emission from cometary parent molecules results from 
radiative excitation by solar infrared radiation followed by 
fluorescence. For the fundamental vibrational bands of water, including 
$\nu_{2}$, emission is not pure resonant fluorescence, as these 
bands are significantly populated by radiative decay from higher 
excited vibrational states. The vibrational fluorescence scheme of 
cometary water is presented by \citet{domi89}. The $\nu_{2}$ band is 
significantly populated by decay of the $\nu_2+\nu_3$ band. The 
resulting emission rate of $\nu_2$ is 2.41 $\times$ 10$^{-4}$ s$^{-1}$ 
at $r_{h} = 1$~AU from the Sun. The $\nu_2+\nu_3-\nu_3$ hot band has an 
emission rate of $7.4 \times 10^{-6}$~s$^{-1}$ and therefore does not 
contribute significantly to the emission observed between 6 and 
7~\micron. Other hot bands (e.g., $(\nu_1+\nu_2+\nu_3)-(\nu_1+\nu_3)$) are 
even weaker. 

We used the model of fluorescence water emission presented by \citet{domi89} 
for analyzing the \spitzer{} data. This model considers five 
excited vibrational states and their subsequent radiative cascades, and 
is an improvement of that presented in detail in \citet{dbm87}, where 
only the $\nu_{2}$ and $\nu_{3}$ bands are considered. Einstein 
coefficients for ro-vibrational transitions are computed using the 2003 
edition of the GEISA spectroscopic database \citep{jacq05}, which includes 
all significant routes leading to $\nu_2$ excitation, including via 
hot bands. Comparison of the line strengths given in 
GEISA with those resulting from the {\it ab-initio} calculations 
of \citet{PS97} verified that the line-by-line relative 
intensities are insensitive (within 3--4\%) to the choice of water line 
lists.

\begin{figure}
\begin{center}
\includegraphics[angle=270,scale=.70]{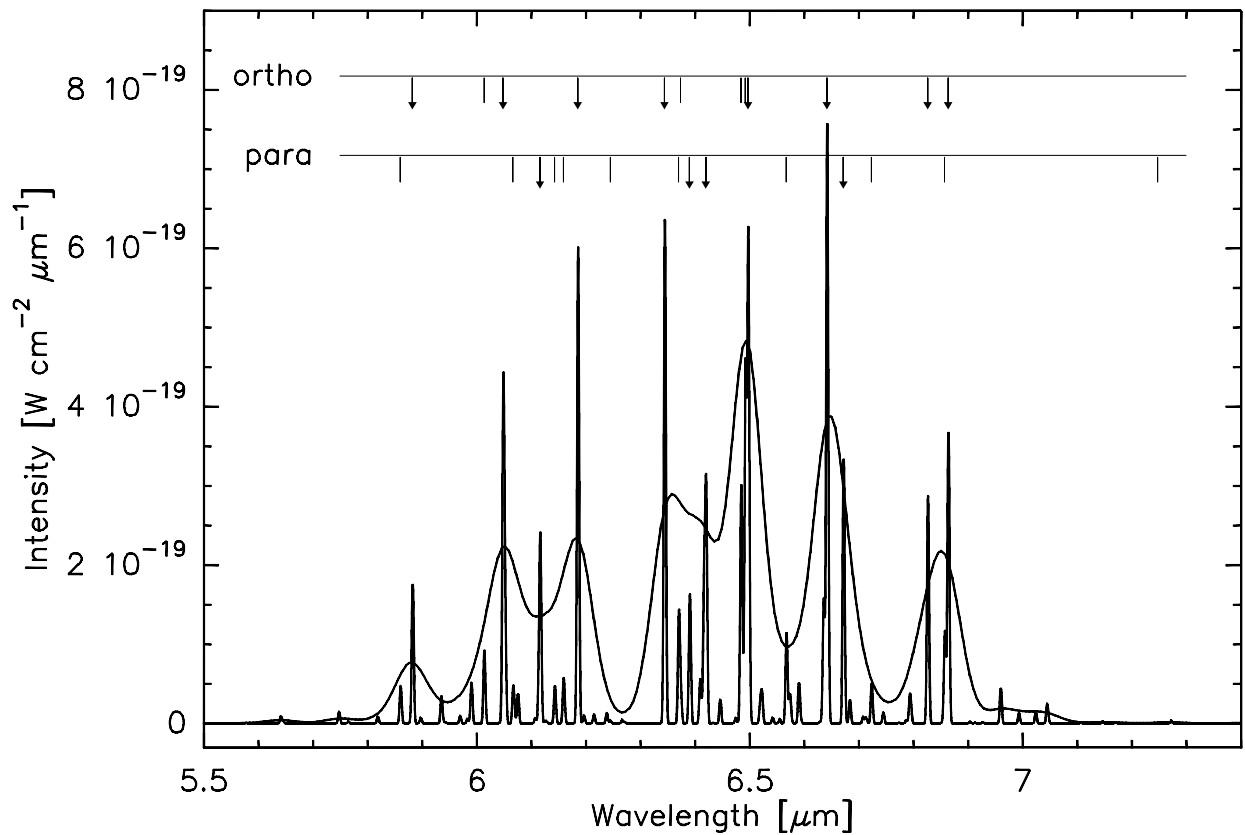}
\end{center}
\caption{Synthetic spectra of the water $\nu_2$ band with 
a high spectral resolution of 0.0044 $\mu m$ (1 cm$^{-1}$)
with the intensity divided by 5, and at the resolution of 
0.065 $\mu m$ corresponding to the \spitzer{} IRS SL2 spectrometer. 
Ortho and para lines are indicated at the top, with the arrows 
showing the strongest lines for both spin species. Calculations pertain to 
a 0.075\arcmin\ field of view radius centered on the nucleus position, 
$Q$(H$_2$O) = 5 $\times$ 10$^{28}$ molec.~s$^{-1}$, 
$r_h$ = 1.760 AU, $\Delta$ = 1.409 AU, $v_{exp}$ = 0.8 km  s$^{-1}$, 
$T_{kin}$ = 40 K, and OPR = 3.
\label{fig:synthe}} 
\end{figure}

\begin{figure}
\begin{center}
\includegraphics[angle=270,scale=.70]{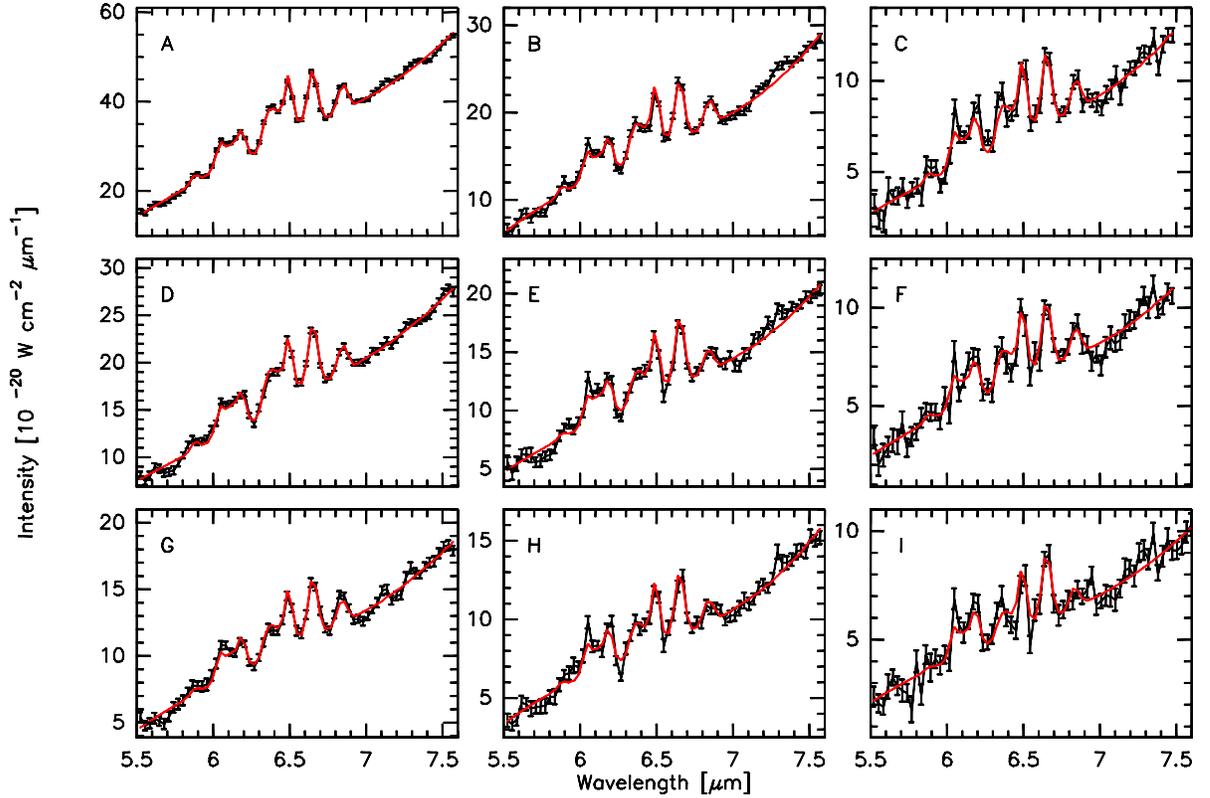}
\end{center}
\caption{Model fits to the spectra of C/2003 K4 (LINEAR) extracted along 
the slit positions (see \S~\ref{obs}, Fig.~\ref{fig:slit}). Capital letters
on the top-left corners correspond to the labels defined in the 
caption of Fig.~\ref{fig:slit}. Data are shown in black with pipeline 
derived errorbars excluding SL2 fringe uncertainty (see text 
\S~\ref{obs-irs}), with the model fits (in red) superimposed. 
Model fitting was performed with the rotational temperature $T_{rot}$ and 
the ortho-to-para ratio taken as free parameters. The derived water 
band intensities are given in Table~\ref{table:tb_k4}. 
Derived $T_{rot}(K)$ and OPRs 
are given in Tables~~\ref{table:tb_k4} and \ref{table:split-fits} 
(see \S~\ref{disc-h20}). The underlying continuum, described by a 
polynomial of degree 5 -- 6, was also fit simultaneously. 
\label{fig:fitcont}} 
\end{figure}

Our water model takes 
into account opacity effects in vibrational excitation and emission, 
using the escape probability formalism. For computing the line-by-line 
fluorescence, we considered 32 ortho and 32 para rotational levels in 
each vibrational state.  The rotational populations in the ground 
vibrational state can be described by a Boltzmann distribution at a 
temperature $T_{rot}$, or the populations can be computed using an 
excitation model that considers the evolving excitation conditions 
experienced by the water molecules as they expand in the coma 
(\S~\ref{trot}). In this detailed model, ro-vibrational line 
intensities are computed for a circular aperture centered on the nucleus. 
We do not expect the results to be significantly sensitive to the shape 
of the aperture, providing the aperture area is conserved. From the model 
output, synthetic \spitzer{} spectra were generated by convolving 
the intensity of the individual ro-vibrational lines with the 
instrumental spectral response of the spectrometer, described by a 
Gaussian. Figure~\ref{fig:synthe} shows examples of synthetic spectra 
obtained for the spectral resolution of SL2 and a 15 times higher 
resolving power. At first glance, the position and relative intensities 
of the peaks in the water modeled spectrum match approximately those  
in the C/2003 K4 (LINEAR) \spitzer{} spectra shown in 
Fig.~\ref{fig:fitcont}, thereby demonstrating that these spectra 
are dominated by water emission.

For fitting the observed spectra, we assumed that the rotational
populations of the ground vibrational state follow a Boltzmann
distribution. The gas expansion velocity, $v_{exp}$, was fixed to
0.8~km~s$^{-1}$. The water photodissociation rate was taken equal to
$1.6 \times 10^{-5}$~s$^{-1}$ ($r_{h} =1$~AU).
This rate takes into account the solar activity at the time of 
the observations following the formalism described by \citet{cro89}. The 
only free parameters of the model are the water production rate $Q$(H$_2$O),
$T_{rot}$, and the OPR.

Overall, opacity effects are small. We computed that they 
affect the total intensity of the
$\nu_{2}$ band by 6\% for the spectrum of C/2003 K4 (LINEAR) acquired 
closest to the nucleus (7.2\arcsec \ offset, Fig.~\ref{fig:slit}). In 
addition, if opacity effects are not properly taken into account 
in the calculations, then the derived OPR also can be 
underestimated (on the order of $\sim 6$\%).

The water band emission is in excess of the dust continuum emission 
(Fig.~\ref{fig:fitcont}). In the first analyses, the 
underlying continuum was determined using a
5-th order polynomial fit, and the residual (continuum-subtracted)
spectra were fit with the water model, applying a least-squares
method that uses the gradient-search algorithm of Marquardt. Continuum
subtraction was not completely satisfactory, as excess continuum
emission remained near 6.26~\micron, while the $\nu_2$ band is almost
free of lines at this wavelength (Fig.~\ref{fig:synthe}). More robust
fits could be obtained by fitting simultaneously the underlying
continuum and the water emission. Thus, we fit the
original spectra with a composite curve consisting of the
modeled water spectrum superimposed on a polynomial. Polynomials of
$5-$ or $6-$degree were used. However, final results were not found to be
significantly sensitive to the choice of the polynomial degree between
3 and 6. 
   
The nominal spectral resolution of SL2 is 0.060~\micron{} near
6~\micron{}. Model fits to the best C/2003 K4 (LINEAR) spectrum 
(7.2\arcsec \ offset) with the spectral resolution 
left as a free parameter yielded $\Delta \lambda = 0.067\pm0.004$~\micron{}, 
agreeing with the edge observation of NGC 7027 (\S~\ref{obs-wave}).  
Results of model fits
given in Table~\ref{table:tb_k4} were obtained with $\Delta \lambda$
fixed to 0.065~\micron{}. However, almost identical results (within
the error) are returned with $\Delta \lambda$ = 0.060~\micron{}. For
example, for the highest signal-to-noise ratio spectrum
(Fig.~\ref{fig:fitcont}, box A), the retrieved OPR is changed from
$2.47\pm0.27$ to $2.31\pm0.24$. We also found that the frequency
calibration in SL2 spectra is likely incorrect by a tenth of the
resolving power. The central wavelengths of $\nu_{2}$ band structures in
C/2003 K4 (LINEAR) spectra are better matched by shifting the 
observed spectra by 0.0032 to 0.0062~\micron, within the errors discussed in
\S~\ref{obs}.  For example, for the 7.2\arcsec\ offset spectrum,
the $\chi^2$ between 5.8 and 7.1~\micron \ is decreased by a factor
2.8 when applying a 0.0062~\micron \ offset. The spectra of comet
C/2003 K4 were shifted by 0.0032 to 0.0062~\micron \ for the model
fits shown in Table~\ref{table:tb_k4} and the corresponding figures
(Figs.~\ref{fig:fitcont} and~\ref{fig:fitk4}).

\subsection{Water Modeling Results\label{disc-h20}}

\begin{figure}
\begin{center}
\includegraphics[angle=270,scale=.70]{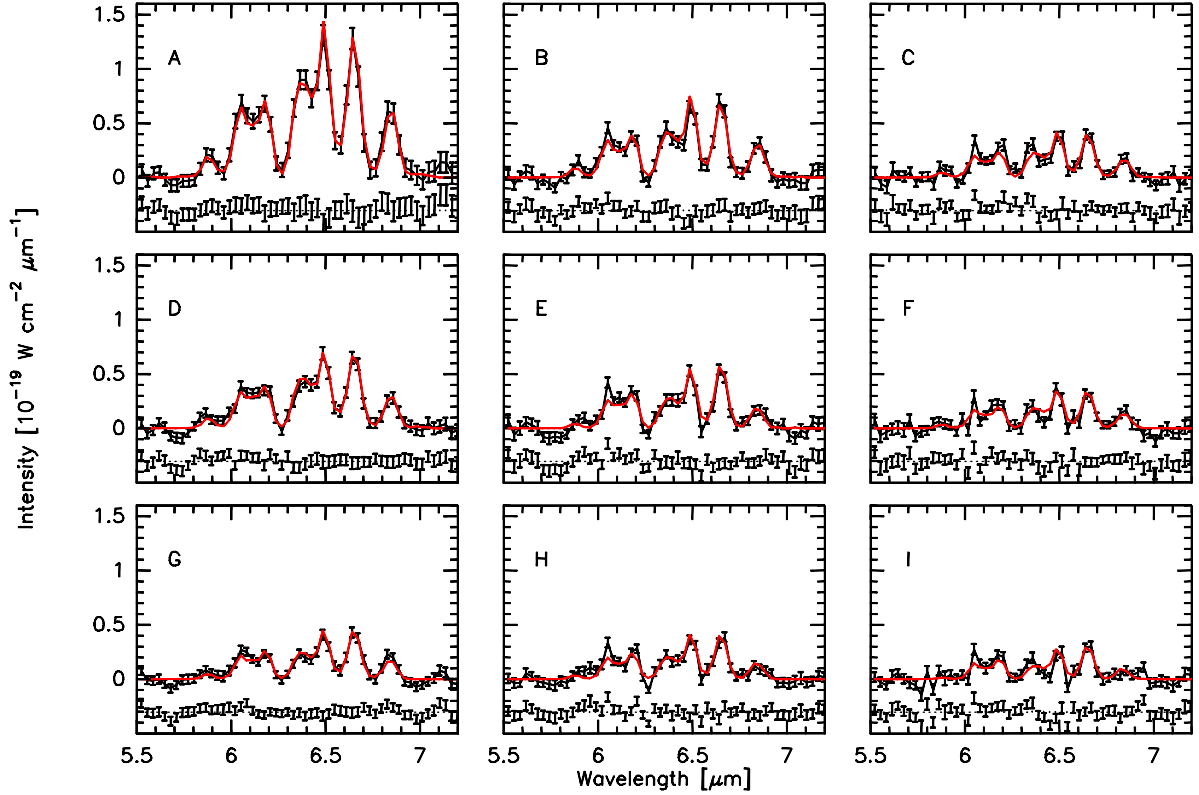}
\end{center}
\caption{Model fits to the spectra of C/2003 K4 
(LINEAR)(see \S~\ref{obs}, Fig.~\ref{fig:slit}). Same as 
Fig.~\ref{fig:fitcont}, with the fitted continuum background 
subtracted. Data are shown in black with errorbars, with the model 
fits (in red) superimposed. Errorbars include here SL2 fringe uncertainty 
(see \S~\ref{obs-irs}). The residuals are shown on the bottom. 
Model fitting was performed with the rotational temperature $T_{rot}$ 
and the ortho-to-para ratio taken as free parameters. The derived 
values are given in Tables~\ref{table:tb_k4} and \ref{table:split-fits} 
(see \S~\ref{disc-h20}). 
\label{fig:fitk4}}
\end{figure}

The best-fit modeled spectra for comet C/2003 K4 (LINEAR) are shown in
Figs.~\ref{fig:fitcont} and~\ref{fig:fitk4}, where in the latter
figure the continuum has been subtracted. Residuals with respect to
observed spectra are shown in the bottom of 
Fig.~\ref{fig:fitk4}. Retrieved model parameters are given in 
Table~\ref{table:tb_k4}. The agreement between
our models and the \spitzer{} spectra is rather 
good for slit extractions A, D, and G, with reduced
$\chi^{2}$ less than 1 (Table~\ref{table:tb_k4}). 

\begin{figure}
\begin{center}
\includegraphics[angle=270,scale=.70]{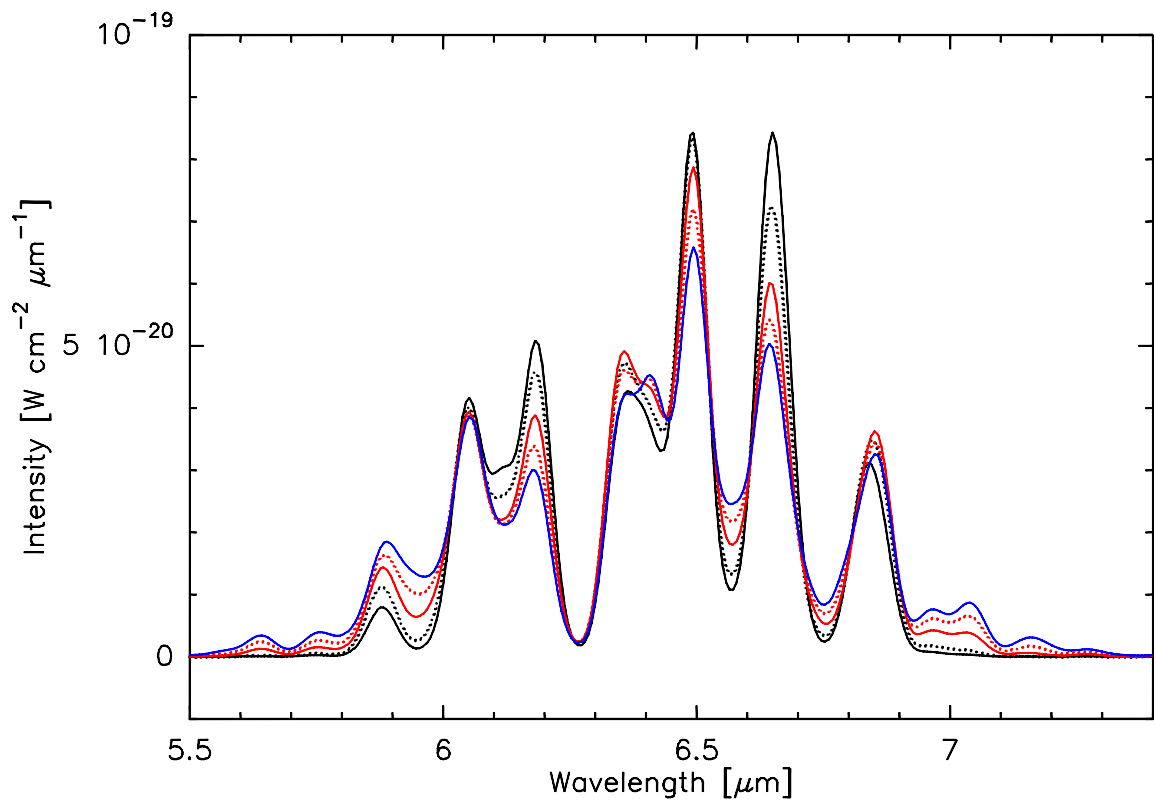}
\end{center}
\caption{Emergent model intensity of the water $\nu_2$ band as a 
function of rotational temperature, $T_{rot}$. The synthetic water 
spectra were generated at the spectral 
resolution corresponding to the \spitzer{} IRS SL2 spectrometer 
(0.065~\micron). Calculations pertain to 
a 0.075\arcmin\ field of view radius centered on the nucleus position, 
$Q$(H$_2$O) = 1 $\times$ 10$^{28}$ molec.~s$^{-1}$, $r_{h}$ = 1.760 AU, 
$\Delta$ = 1.409 AU, $v_{exp}$ = 0.8 km  s$^{-1}$, with $T_{rot}$ 
equal to 20~K ({\it solid black line}), 30~K ({\it dotted black 
line}), 50~K ({\it solid red line}), 70~K ({\it dotted red line}), and 
90~K ({\it solid blue line}).  
\label{fig:temp_trot_var}}
\end{figure}

Some excess 
emission ($>$~3-$\sigma$ deviation) is noticeable at 6.05~\micron \ in 
most of the spectra at offset $>$~18\arcsec. 
Models that incorporate values of $T_{rot}$ higher than those 
determined from model fitting (Table~\ref{table:tb_k4}) reduce the
residual continuum emission near 6.05~\micron \ but are inconsistent
with the relative 
water line intensities arising from intrinsically stronger lines
(Fig.~\ref{fig:synthe}) measured longward 6.3~\micron.
The variation in emergent water 
line intensities for $20 < T_{rot}(\rm{K}) < 90$ which produce the 
broad emission feature from 5.5 to 7.0~\micron \ when the models 
are convolved to the resolution of the \spitzer{} IRS SL2 are 
shown in Fig.~\ref{fig:temp_trot_var}.
The 6.05~\micron \ peak
(mainly ortho $2_{12}$--1$_{01}$ $\nu_2$ line) is more intense than
the 6.18~\micron \ (mainly $1_{10}$--1$_{01}$ line) peak 
only for high $T_{rot}$ ($>$ 60~K). At low
$T_{rot}$, these two lines result essentially from IR pumping from the
1$_{01}$ ground state rotational level: the ratio of their intensities
$I$(6.18 $\mu$m)/ $I$(6.05 $\mu$m) then depends uniquely on
ro-vibrational Einstein $A$-coefficients and is predicted to be 1.5.

Weak residual emission (2-$\sigma$ deviation) is also observed at 
5.90--5.95~\micron\ (see B, E, H extractions in Fig.~\ref{fig:fitcont}). 
This excess emission does not coincide in wavelength to the 
position of the 5.88~\micron \ peak of the water band, and may be
related to flaws in background subtraction. 

Our derived values of $T_{rot}$ and OPR could be affected by the
residual emission present between 5.90--6.1~\micron. This emission 
(in excess of fringe artifact contamination) may arise from 
sources of weak line and continuum emission other than water 
not accounted for in our models. To quantify such effects, we 
derived estimates of $T_{rot}$ and OPR 
by independently fitting two partial subsections of the spectra 
(5.85--6.3~\micron\ and 6.3--7.0~\micron) and 
any continua shortward of 5.8~\micron \ and longward of 7.0~\micron\ 
for the brightest slit extractions. These independent model fits, 
including those derived from the best reduced $\chi$-squared fit of the 
entire spectra, are summarized in Table~\ref{table:split-fits}.
Consistent OPR values are obtained from the independent model fits. 
The derived $T_{rot}$ values are higher as a result of the
5.90--6.1~\micron{} excess flux only when the 5.85--6.3~\micron\ part of
the spectrum is considered. Table~\ref{table:tb_k4} provides model 
parameters for all slit extractions retrieved by fitting the 
6.3--7.0~\micron~part of the water spectrum. However, these 
latter fit parameters are not significantly different from those 
derived by fitting the entire spectrum.   

\subsection{Rotational Temperature\label{trot}}

\begin{figure}
\begin{center}
\includegraphics[angle=270,scale=.50]{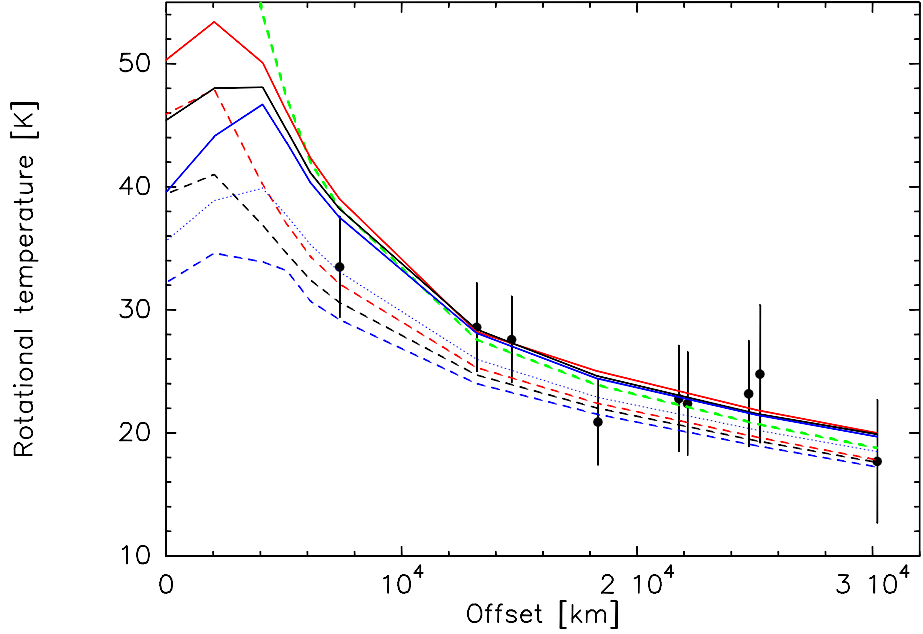}
\end{center}
\caption{Rotational temperature as a function of offset with respect to 
nucleus position in comet C/2003 K4 (LINEAR). Data points (black 
dots with errorbars) correspond to 
the slit extractions shown in Fig~\ref{fig:slit}. Curves show the 
rotation temperature extracted from synthetic $\nu_2$ spectra computed 
with a kinetic temperature $T_{\rm kin}$ of 30, 40, 50, 
and 100~K ({\it blue}, {\it black}, {\it red} and {\it green} lines, 
respectively) and different models of the electron density 
(described by the $x_{ne}$ parameter): $x_{ne}$ = 1 (plain lines),  
$x_{ne}$ = 0.5 (dotted lines), $x_{ne}$ = 0.2 (dashed lines). For
clarity, results obtained with $x_{ne}$ = 0.5 are only plotted for 
$T_{\rm kin}$ = 30 K; for $T_{\rm kin}$ = 100~K, we only show the model
output corresponding to $x_{ne}$ = 0.2.  
\label{fig:k4trot}}       
\end{figure}

The inferred rotational temperatures are between 18 and 34~K 
and tend to decrease with increasing distance from the nucleus 
(Fig.~\ref{fig:k4trot}). The rotational temperatures  
are not constant with respect to cometocentric 
distance out to $3 \times 10^{4}$~km within the 
uncertainties. A linear least squares fit to 
$T_{rot}$ in Table~\ref{table:tb_k4}, as a function of distance to 
nucleus, gives a slope of $-(6.1\pm2.2)$ $\times$ 10$^{-4}$~K~km$^{-1}$.   
Such behavior is expected. In the inner coma, 
collisions are important and the rotational levels of the fundamental 
vibrational state are thermalized at the local temperature. However, in 
the outer coma radiative pumping prevails and the rotational population 
of the water molecules reaches a cold fluorescence equilibrium 
\citep[e.g.,][]{dbm87}. How the populations evolve from thermal to 
fluorescence equilibrium depends of the density of the collisional 
partners, and related collisional cross-sections.  
Other attempts to examine the variation in water rotational 
temperature with cometocentric distance \citep{bonev07} are restricted to 
the inner (\ltsimeq $10^{3}$~km) collision-dominated coma.  

Water excitation models currently developed by various investigators 
include both H$_2$O-H$_2$O and H$_2$O-e$^-$ 
collisions \citep{biv97,ben04,zak07}. For our analysis, we use the excitation 
model of \citet{biv97}, which differs from the 
models of \citet{ben04} and \citet{zak07} by the method used to solve 
radiation trapping effects, but yields almost similar results \citep{zak07}. 
The electron density and radial temperature distribution is based on the 
measurements of 1P/Halley made by the Giotto mass 
spectrometers \citep[e.g.,][]{eberhardt95}, to which scaling 
factors are applied to account for variations with water production rate 
and heliocentric distance. The parameter $x_{ne}$ is a multiplying factor 
to the electron density, normalized to the 1P/Halley Giotto measurements 
($x_{ne} (\rm{1P/Halley}) = 1$). The output of the model is 
the rotational populations in the ground vibrational state as a function of 
distance to nucleus. The populations are included 
in the water infrared fluorescence model to simulate water spectra at 
offset positions. For direct comparison 
with the observations, rotational temperatures are derived by fitting 
the synthetic spectra, as was done for the observed spectra.    

Figure~\ref{fig:k4trot} shows the evolution of 
$T_{rot}$ predicted for $x_{ne}$ values of 0.2, 0.5, and 1.0, 
and inner coma kinetic 
temperatures $T_{\rm kin}$ of 30, 40, 50, and 100~K.  These values of the 
$x_{ne}$ parameter were selected because the electron density is rather 
uncertain and mapping of the 557 GHz H$_2$O line favors $x_{ne}$ $\sim$ 0.2 
\citep{biv07}. The predicted increase in $T_{rot}$  
at $\sim$ 2000 km offset is due to thermal excitation by hot electrons. 
In the electron density model, this distance corresponds to the 
contact surface, $R_{CS}$,  where the electron temperature and density 
undergo a steep increase \citep{xie92}. Beyond $\sim$ 2$R_{CS}$, 
$T_{rot}$ decreases because the effect of excitation by electronic 
collisions become less efficient with respect to radiative decay and
the fluorescence equilibrium of the ground vibrational state is cold. 
In the 7000--15000 km region, the model predicts a decrease in $T_{rot}$ of 
$\Delta T_{rot}$ = 6 to 11~K for the considered parameters, in 
contrast to the observed decrease of $\sim$ 5 to 6~K.  If one considers 
uncertainties associated with $T_{rot}$ derived from the \spitzer{} 
spectra, models with $T_{\rm kin} = 30-90$~K and $x_{ne} = 0.2-0.5$ are 
satisfactory. A $x_{ne}$ value of 1.0 does not fit the data obtained 
at 7.2\arcsec \ offset (Fig.~\ref{fig:k4trot}), in agreement 
with \citet{biv07}. The kinetic temperature is poorly constrained 
at the sampled cometocentric distances because, $T_{rot}$ retains
little memory of excitation conditions prevailing in the collisional
region.        

\subsection{Ortho-to-Para Ratio\label{otpr}}

\begin{figure}
\begin{center}
\includegraphics[angle=270,scale=.750]{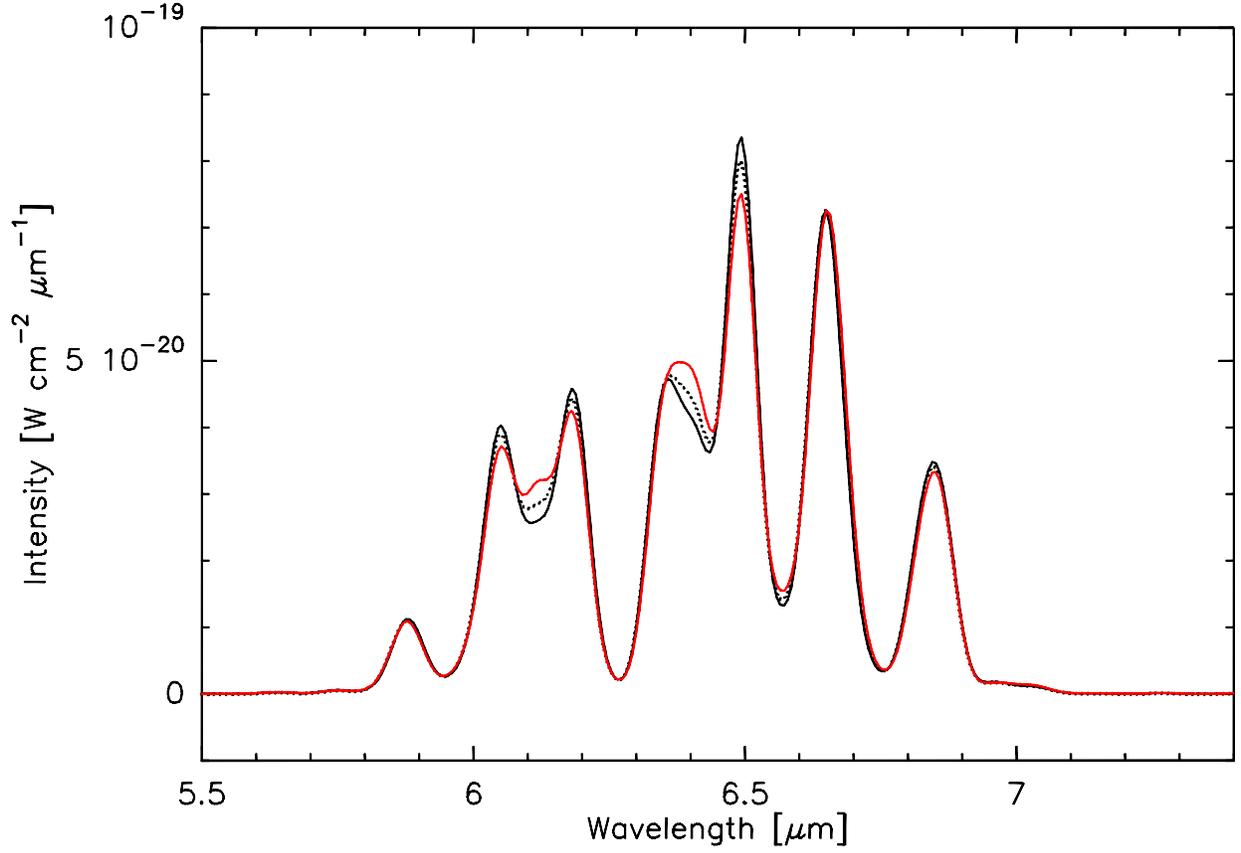}
\end{center}
\caption{Emergent model intensity of the water $\nu_2$ band as a 
function of the ortho-to-para (OPR) ratio. OPRs of 
2.0 ({\it solid red line}), 2.5 ({\it dotted black line}), and 
3.0 ({\it solid black line}) are 
depicted. The synthetic water spectra were generated at 
the spectral resolution corresponding to the \spitzer{} IRS SL2 
spectrometer (0.065~\micron). Calculations pertain to 
a 0.075\arcmin\ field of view radius centered on the nucleus position, 
$Q$(H$_2$O) = 1 $\times$ 10$^{28}$ molec.~s$^{-1}$, $r_{h}$ = 1.760 AU, 
$\Delta$ = 1.409 AU, $v_{exp}$ = 0.8 km  s$^{-1}$, and $T_{rot} = 30$~K.
\label{fig:opr_2to3}}       
\end{figure}

Though the spectral resolution of the SL2 spectra is low ($R \approx
100$), model fitting provides an accurate measurement of the OPR in
comet C/2003 K4 (LINEAR) for the high signal-to-noise spectra
(Tables~\ref{table:tb_k4} and \ref{table:split-fits}).  The OPR 
can be retrieved from $\nu_{2}$ band
spectra obtained at low resolving power because several para lines are
well separated in wavelength from 
strong ortho lines (Fig.~\ref{fig:synthe}). 
The band regions most sensitive to the OPR lie at 6.12 and 6.4~\micron.
At 6.12~\micron, emission is dominated by the $1_{11}$--$0_{00}$
para line, and three other significant para lines
(Fig.~\ref{fig:synthe}). Since the nearby 6.05 and 6.18~\micron \ peaks 
are mainly due to ortho lines ($2_{12}$--$1_{01}$ and $1_{10}$--$1_{01}$, 
respectively), the intensity ratios $I$(6.12$\mu m$)/$I$(6.18$\mu m$)
and $I$(6.12$\mu m$)/$I$(6.05$\mu m$) increase with decreasing
OPR. Similarly, at 6.4~\micron, the
contributions from para lines ($0_{00}$--$1_{11}$ and $1_{11}$--$2_{02}$)
dominate the spectrum and the intensity 
ratio $I$(6.4$\mu m$)/$I$(6.5$\mu m$) is a function of the OPR. 
The variation in water $\nu_{2}$ band features with OPR, 
for $T_{rot} = 30$~K, at the \spitzer{} IRS spectral resolution is 
illustrated in Fig.~\ref{fig:opr_2to3}. From the models depicted in 
this figure, the peak intensities of the 6.05~\micron, 6.18~\micron, 
and the 6.50~\micron \ features increase by $\approx 8$\%, $\approx 8$\%, 
and $\approx 13$\% respectively as the OPR changes from values of 
2 to 3. The intensity at 6.12 and 6.4~\micron \ decreases 
in turn by $\approx 18$\%. However, the 6.64 and 6.85~\micron \ 
water features remain constant and their intensity ratio  
only depends upon $T_{rot}$. The different behaviors of 
the intensity of the features with $T_{rot}$ and OPR make the 
accurate measurement of these two parameters possible.

For the comet C/2003 K4 (LINEAR) spectrum at position A, 
7.2\arcsec \ offset from the nucleus position (Fig.~\ref{fig:slit}), we 
obtain an OPR = $2.47\pm0.27$ when fitting the entire 
water spectrum (Table~\ref{table:split-fits}). The other spectra yield OPRs 
consistent with this value (Tables~\ref{table:tb_k4}, \ref{table:split-fits}). 
Conversions between ortho to para states by radiative transitions 
or by collisions in the coma have very low probability. The constancy 
of the OPR in the coma also has been convincingly demonstrated in comet
C/2004 Q2 (Machholz) by \citet{bonev07}. The weighted mean 
of all OPR values in comet C/2003 K4 (LINEAR) given in 
Table~\ref{table:split-fits} (fits to 5.8--7.0~\micron\ region) 
is $2.43\pm0.15$. The OPR value derived for the aperture slit 
closest to the nucleus (labeled A), $2.47\pm0.27$, corresponds to a 
spin temperature $T_{spin} = 28.5^{+6.5}_{-3.5}$~K. 

The reduced $\chi^2$ between 5.8 and 7.1~\micron \  
obtained for the spectrum at 7.2\arcsec \ offset is 0.5. When the OPR 
is fixed to OPR = 3, the reduced $\chi^2$ is $\simeq 15$\% higher 
(45\% higher when fitting 6.3--7~\micron \ partial spectrum).
Figure~\ref{fig:k4opr} shows the model fit obtained in this case which 
yields $T_{rot} = 30.5\pm3.3$~K, a value close to that obtained with OPR = 
2.47 (Table~\ref{table:split-fits}). There is significantly higher discrepancy 
between 6.3 and 6.4~\micron\mbox{} in the two models.

The OPR (and $T_{rot}$) determination relies on the assumption that 
water emission dominates the 5.8--7.2~\micron \  C/2003 K4 (LINEAR) spectrum.  
Misleading results can be obtained when extra emission from other
constituents is present. In addition to the 6~\micron \ residual emission
discussed in \S~\ref{disc-h20}, PAH emission 
peaks \citep[e.g.,][]{pee02} near 6.2~\micron, i.e., in the region where 
the shape of H$_2$O band depends on the OPR (Fig.~\ref{fig:k4opr}). 
Because the independent model fits of 
partial subsections of the spectra provide consistent OPR 
values (Table~\ref{table:split-fits}), our OPR determinations
are likely not affected by unaccounted species.   
We also conclude that the dominant source of emission in the 
SL2 wavelength regime for comet C/2003 K4 (LINEAR) is the 
water $\nu_{2}$ band. 

Our derived water OPR for C/2003 K4 (LINEAR) is comparable to 
to values derived for Oort Cloud (nearly-isotropic) comets such as 
C/1995 O1 (Hale-Bopp), C/1999 H1 (Lee), C/1999 S4 (LINEAR),
C/2001 Q1 (NEAT), and C/2001 A2 (LINEAR), although lower 
($\sim$ 15-20\%) than that reported for Jupiter-family (ecliptic) comets, 
for example 103P/Hartley~2 or 1P/Halley \citep{crovs00, del05, kawakita06}
or C/2004 Q2 (Machholz) \citep{bonev07}. The value of 
$T_{spin} = 28.5^{+6.5}_{-3.5}$~K for C/2003 K4 (LINEAR) 
is suggestive of precometary ice formation in a cold molecular cloud 
environment devoid of secondary processing in a warm solar 
nebula \citep{kawakita06}, although 
the precise interpretation of the OPR as a probe of the primordial 
formation zones of comets in the protosolar nebula remains 
vexing \citep{crovs07}. We also do not have spectroscopic measurements of 
other common cometary ices such as ammonia or methane in coma of 
C/2003 K4 (LINEAR).  Thus we are unable to ascertain whether the 
ices incorporated into the nucleus of C/2003 K4 (LINEAR) share the same 
chemical composition and homogeneity of $T_{spin}$ as that found 
for other comets of diverse dynamical classes \citep{crovs07}.

\subsection{Water Production Rate\label{h20rate}}

The intensity of the $\nu_2$ band measured for C/2003 K4 along the 
nine slit extractions is given in Table~\ref{table:tb_k4}, and is plotted 
as a function of offset in Fig.~\ref{fig:modflux}. The evolution with 
distance to nucleus is consistent with that computed using a Haser 
distribution for the water density and $Q$(H$_2$O) = $(2.43\pm~0.25) \times 
10^{29}$~molec.~s$^{-1}$, where the error includes a 10\% uncertainty 
in the IRS absolute calibration \citep{irsdh}. Some deviations are 
observed, which may be related to asymmetries in the density distribution 
and/or (for the noisy spectra) incorrect background subtraction.

The derived pre-perihelion ($r_{h}$ = 1.76 AU) water production rate is 
consistent with OH 18-cm observations performed with the Nan\c{c}ay 
radio telescope which yield 
$Q$(H$_2$O) $\sim$ 2 $\times$ 10$^{29}$ molec.~s$^{-1}$ 
at the epoch of the {\it Spitzer} observations 
(Crovisier et al., personal communication). Post-perihelion 
measurements obtained from H$_2$O 557 GHz line observations using the
Odin satellite give a $Q$(H$_2$O) about a factor of two lower 
at $r_{h} = 1.7 - 1.8$~AU \citep{biv07}, which suggests a pre/post-perihelion 
asymmetry in the gaseous activity of the comet. Similar asymmetrical 
perihelion production rates of water and other volatiles has been 
observed in other comets, including C/1995 O1 (Hale-Bopp) 
\citep[e.g., Fig.~3 of][]{biver-hb97} and possibly 1P/Halley
\citep[e.g., Fig.~6 of][]{gehrzet05}.

\section{Residual Emission Between 5.5 and 7.6~\micron{}\label{residuals}}

\begin{figure}
\begin{center}
\includegraphics[angle=270,scale=.70]{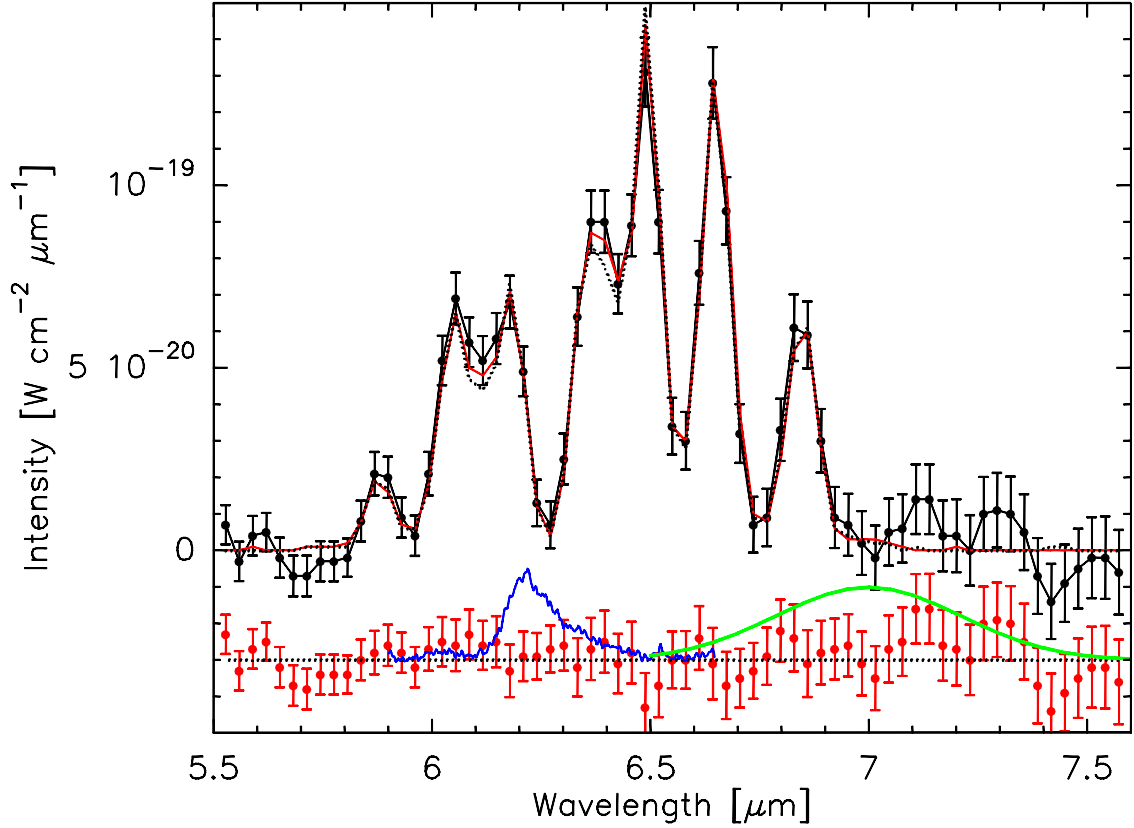}
\end{center}
\caption{Model fits to the spectrum of C/2003 K4 (LINEAR) at 7.2\arcsec 
\ offset from the nucleus.  Data are shown in black dots with errorbars
that include noise from  spectral fringing (see \S~\ref{obs-irs}).
Synthetic spectra with OPR = 2.47 and OPR = 3 are shown in red and in 
black dotted line, respectively. The former fit was obtained with the 
OPR set as a free parameter, in contrast to the second fit. The retrieved 
$T_{rot}$ are $30.7\pm3.2$~K and $30.5\pm3.3$~K, respectively. The
residual spectrum from the former fit is shown in red on the bottom, 
on which are superimposed in arbitrary units an interstellar PAH spectrum 
typical of class A sources from \citet{pee02} ({\it blue} spectrum), and 
a model of carbonate emission from \citet{lisse06} ({\it green} spectrum). 
\label{fig:k4opr}}       
\end{figure}

\begin{figure}
\begin{center}
\includegraphics[angle=270,scale=.50]{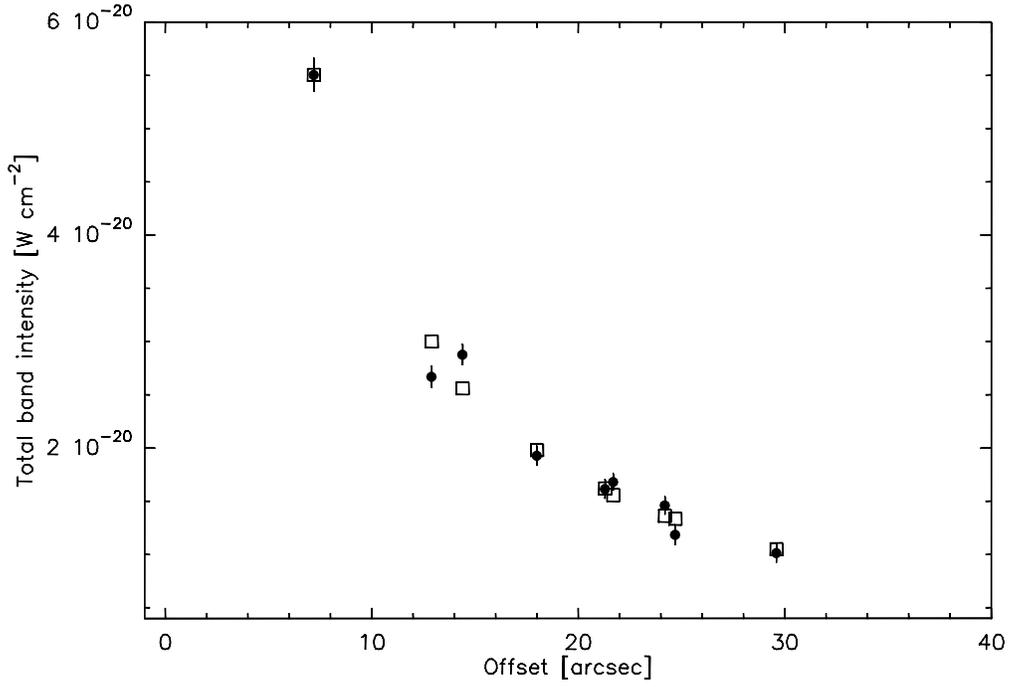}
\end{center}
\caption{Total band intensity between 5.8 and 7~\micron \ in C/2003 K4 
(LINEAR) as a function of offset with respect to nucleus position. Data 
points (black dots with errorbars) correspond to the slit extractions 
shown in Fig~\ref{fig:slit}. Squares show expected water $\nu_2$ band 
intensities for $Q(\rm{H}_{2}\rm{O}) = 2.43 \times 
10^{29}$~molec.~s$^{-1}$, $v_{exp}$ = 0.8 km s$^{-1}$ and 
$\beta$(H$_2$O) = 1.6 $\times$ 10$^{- 5}$ s$^{-1}$ at $r_h$ = 1 AU. 
\label{fig:modflux}} 
\end{figure}

\citet{lisse06} report emission from carbonate minerals at
6.5--7.2~\micron{} and organic (PAH) emission at 6.2~\micron{} in the
spectrum of comet 9P/Tempel after collision with the
\textit{Deep~Impact} impactor. In a subsequent paper, \citet{lisse07}
claim the detection of these emission features in the ISO spectrum of 
comet C/1995 O1 (Hale-Bopp) published by \citet{cro97b}. However, a 
re-analysis of the ISO observations of comet Hale-Bopp by \citet{cro07} 
does not confirm the detection of PAHs reported by \citet{lisse07}.
Furthermore, \citet{cro07} demonstrate that carbonate
emission at 7~\micron{}, though possibly marginally present,
is fainter by a factor 2 to 3 than asserted by \citet{lisse07}.
Figure~\ref{fig:k4opr} shows representative spectra of 
PAHs and carbonate minerals compared to the best \spitzer{} 
spectrum of comet C/2004 K4 (LINEAR). Our synthetic water 
spectrum wholly accounts for any emission in-excess of the continuum at the 
wavelengths of PAHs and carbonate emission. As 
discussed in \S~\ref{disc-mfit} and \S~\ref{disc-h20}, residual emission is 
only marginally present in some spectra near 5.9 and 6.05~\micron{} in 
various apertures in the C/2003 K4 (LINEAR). Since PAH features are 
narrow ($\Delta \lambda \sim$ 0.15~\micron) and peak near the water 
6.18~\micron~pattern, a significant contribution of PAHs in the spectrum 
would have resulted in an intensity ratio 
$I$(6.18 $\mu$m)/$I$(6.05 $\mu$m) higher than observed. Similarly,   
carbonate emission, if present, would have been seen directly on the
original spectra (continuum background included, Fig.~\ref{fig:fitcont}) 
longward of 7~\micron. Indeed the 3-$\sigma$ upper limit to
the peak intensity of any carbonate or PAH emission, computed
from the residual emission between 6.15 to 6.30~\micron{}
and 6.75 to 7.25~\micron \ using the representative PAH and 
carbonate spectra shown in Fig.~\ref{fig:k4opr}, does not
exceed 7 to $8 \times 10^{-21}$~W~cm$^{-2}$~\micron$^{-1}$ 
(\ltsimeq $10^{-3}$~Jy) in the C/2003 K4 (LINEAR) spectrum 
at 7.2\arcsec \ offset position. 

\begin{figure}
\begin{center}
\includegraphics[angle=0,scale=0.75]{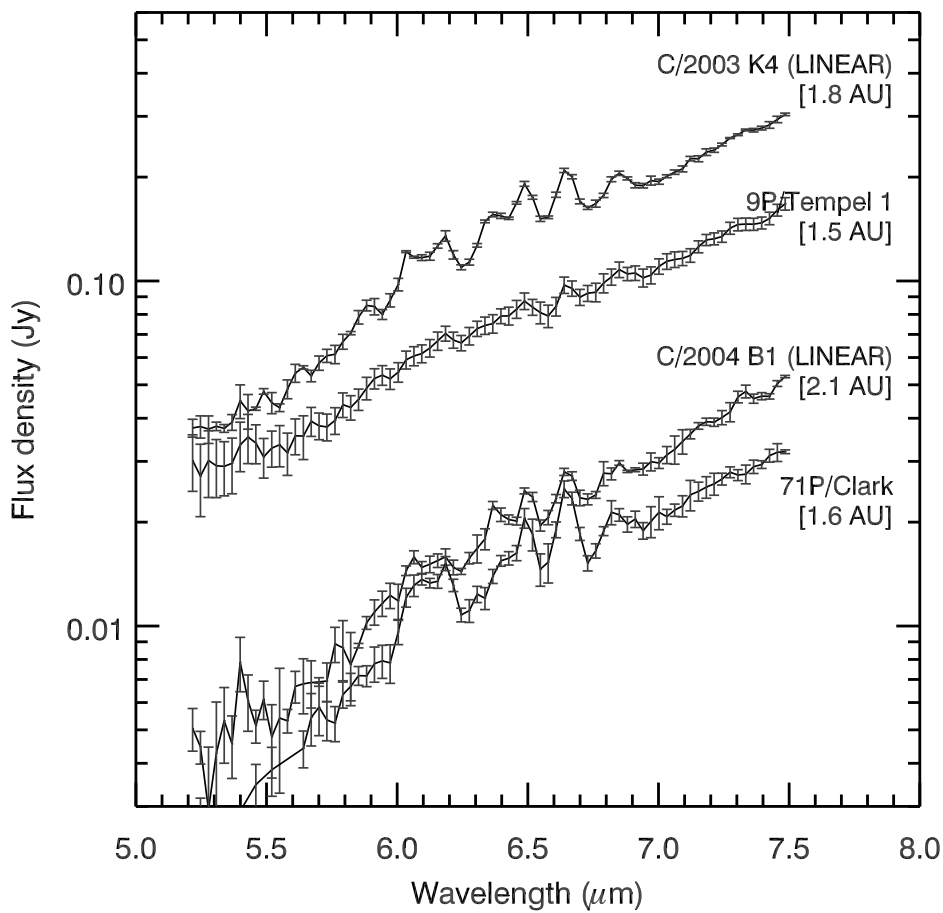}
\end{center}
\caption{A comparison of the $\nu_{2}$ water bands in
Oort Cloud comets C/2003 K4 (LINEAR), C/2004 B1 (LINEAR), and Jupiter-family
comets 71P/Clark and 9P/Tempel~1 detected with the \spitzer{} IRS. The
data have not been scaled or offset in flux with respect to each other
and the spectra do not have the continuum emission removed.
Our synthetic model spectrum suggests that water dominates
emission in-excess of the continuum at wavelengths between 5.7 
and 7~\micron. 
\label{fig:h20_all}}
\end{figure}

As depicted in Fig.~\ref{fig:h20_all}, \spitzer{} spectra of  
comets C/2004 B1 (LINEAR), 71P/Clark, and 9P/Tempel~1 show evidence 
for $\nu_{2}$ water emission arising from sublimating 
ices in their comae. A detailed study, similar to that 
presented in this paper, is required to investigate whether 
emission from other compounds is present in these 
spectra \citep{woodward07}. 
   
\begin{figure}
\includegraphics[angle=270,scale=0.60]{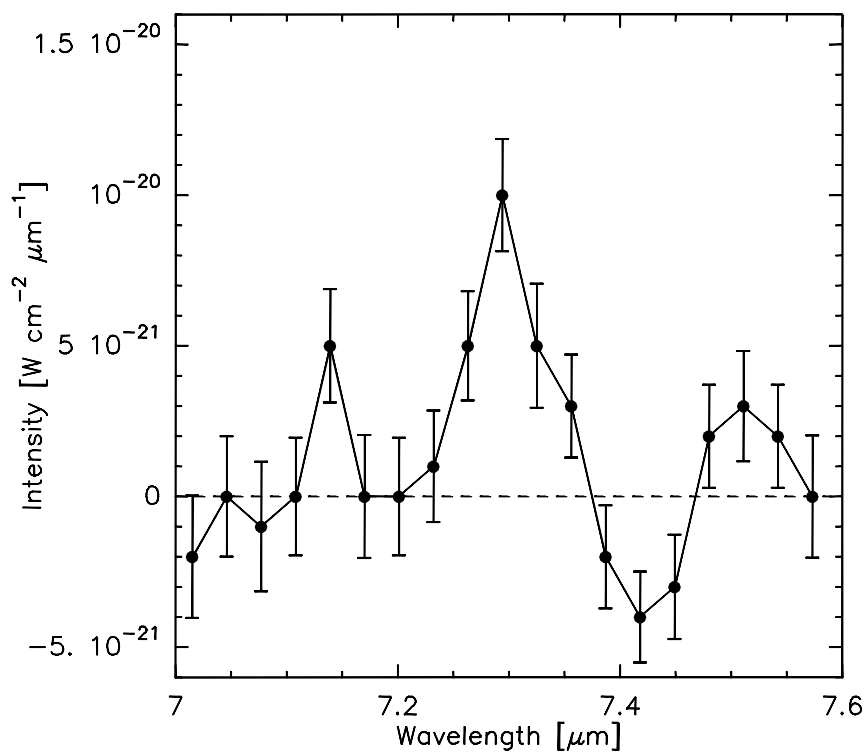}
\caption{Background-subtracted 7--7.6~\micron\ spectrum of 
C/2003 K4 (LINEAR) showing the detection of a weak emission feature 
at 7.3~\micron . This spectrum has been obtained by averaging slit 
extractions 0-27, 0-22, 1-27, and 1-22. The uncertainty plotted are
the error from extractions of the pipeline reduced individual BCDs and do
not include any small excess noise arising from spectral fringing (see
\S~\ref{obs-irs}).
\label{fig:7p3mu-spec}}
\end{figure}

From Fig.~\ref{fig:k4opr}, we see that a narrow 
($\Delta \lambda \sim $ 0.1~\micron{}) residual emission feature 
is present near $\sim$7.3~\micron~(1370 cm$^{-1}$).
A small spectral segment of the \spitzer{} IRS data near 7.3~\micron, 
shown in Fig.~\ref{fig:7p3mu-spec}, provides detailed, close-up 
view of this emission feature. The origin of this feature is
unclear. The peak flux and integrated feature flux is in 
excess of that anticipated from spurious fringe signal power. A 
possible candidate 
is emission from the SO$_2$ \ $\nu_{3}$ band at 7.34~\micron \ that 
has a fluorescence emission rate at 1 AU from the Sun of 
$6.6~\times~10^{-4}$~s$^{-1}$ \citep{cro02}. Synthetic spectra of 
SO$_2$\ $\nu_{3}$ band obtained using the HITRAN database \citep{roth05} 
approximately match the width of the feature, but the central 
wavelengths do not coincide. In addition, the measured intensity 
in the spectrum obtained at 7.2\arcsec \ offset in the coma of 
comet C/2003 K4 (LINEAR), $\simeq 10^{-21}$~W~cm$^{-2}$, would 
imply a SO$_2$/H$_2$O production rate 
ratio of 2.5\%, a factor 10 times higher
than measured in comet C/1995 O1 (Hale-Bopp) \citep{bock00}. Therefore, 
it seems unlikely that the observed 
7.3~\micron \ feature is due to SO$_2$. The NIST 
database\footnote{http://webbook.nist.gov/chemistry/} provides band 
positions of a number of gas phase species, including organics. No 
satisfactory candidate could be 
found. For example, methyl formate HCOOCH$_3$, identified in 
cometary atmospheres \citep{bock00}, 
has a band of medium strength at 1371 cm$^{-1}$, but also   
a much stronger band at 1754 cm$^{-1}$ (5.7~\micron) which is not seen 
in the \spitzer{} spectra. Acetic acid exhibits a strong band 
at 1375 cm$^{-1}$, but a still stronger one at 1248 cm$^{-1}$ 
(6.94~\micron). If the feature is originating from a gas phase species, 
then the abundance of this molecule relative to water should be on 
the order of 1\% or more, based on the measured intensity and typical 
fluorescence emission rates in cometary environment.        

The 7.3~\micron\ wavelength corresponds to the characteristic vibrational 
frequency of the CH$_3$ ``umbrella'' deformation 
mode ($\sim$1375 cm$^{-1}$). A 7.3~\micron{} absorption feature 
has been detected in some galactic and 
extragalactic sources \citep[e.g.,][]{chi00,spoo00}, and 
assigned to aliphatic hydrocarbons. However, an absorption signature 
at 6.85~\micron \ is also observed, corresponding to CH$_2$ bending 
vibrations. Spectra of various carbonaceous 
refractory materials, including chondritic material, that contain aliphatic 
chains show that these two features are present as a doublet with 
intensity ratio $I(6.8 \mu$m$)$/$I(7.3 \mu$m$)$ $>$ 1 \citep{pen02}. 
In contrast, no residual emission is observed at 6.8~\micron \ in 
the comet C/2003 K4 (LINEAR) spectrum. Therefore, aliphatic hydrocarbons
are not likely the source of the cometary 7.3~\micron~feature, 
although such compounds have been identified in the material captured from 
comet 81P/Wild~2 by the Stardust spacecraft \citep{kel06}.         

A few galactic and extragalactic sources exhibit a weak emission feature 
at 7.3--7.4~\micron~(with no 6.85~\micron \ counterpart)   
that shows up on the wing of the well known strong 7.7~\micron\ complex  
attributed to CC stretching/CH in plane bending vibrations of aromatics 
(likely PAHs) compounds \citep{pee02}. Based 
on theoretical calculations of expected CC stretching band positions of 
PAHs of various complexity \citep[see][]{pee02}, this weak component 
is likely a PAH signature. However, the comet feature may have a different
origin as no strong 6.8 and 7.7~\micron \ PAH emission is evident in the
\spitzer{} spectra of C/2003 K4 (LINEAR).

\section{CONCLUSIONS\label{concl}}

We have observed the $\nu_{2}$ vibrational band of water in 
comet C/2003 K4 (LINEAR) within 5.5 to 7.6~\micron \ \spitzer{} IRS
spectra, deriving a water production rate of $(2.43\pm0.25) \times 
10^{29}$~molec.~s$^{-1}$ when the comet was at a pre-perihelion 
heliocentric distance of 1.760~AU.
Although the IRS spectra are of moderate resolution, modeling of
the observed emission in the 5.7 to 6.8~\micron \ region constrained
the water spin temperature to be $28.5^{+6.5}_{-3.5}$~K.
The measured $T_{spin}$ is comparable to that of 
other Oort cloud comets and suggestive of a common formation zone for the 
precometary water ices that eventually agglomerated into the nuclei, 
though the precise interpretation of the OPR as a probe of the primordial 
formation zones of comets in the protosolar nebula remains 
controversial \citep{crovs07}. The observed decrease (at 3-$\sigma$ 
confidence level) of the water rotational temperature with 
cometocentric distance is compatible with evolution from thermal
to fluorescence equilibrium and constrains somewhat the role of
electron collisions in water excitation. The kinetic temperature 
of the gas is poorly constrained.    

Neither emission from carbonates nor PAHs was necessary to account
for any emission in excess of the continuum at wavelengths between
5 to 7~\micron, suggesting that these species are not present in the 
coma of C/2003 K4 (LINEAR) at the abundance levels measured by 
\citet{lisse06} in comet 9P/Tempel 1. However, an emission feature at 
$\sim7.3$~\micron \ is observed that remains unidentified, as potential
emission candidates, the SO$_{2}$ \ $\nu_{3}$ band or CH$_{3}$
deformation modes, can be discounted.

\acknowledgements

This work is based on observations made with the {\it Spitzer} Space
Telescope, which is operated by the Jet Propulsion Laboratory,
California Institute of Technology under a contract with NASA. Support
for this work was provided by NASA through an award issued by
JPL/Caltech.  Support for this work also was provided by NASA through
contracts 1263741, 1256406, and 1215746 issued by JPL/Caltech to 
the University of Minnesota.  C.E.W. and M.S.K. acknowledge support 
from the National Science Foundation grant AST-037446.  M.S.K. acknowledges 
support from the University of Minnesota Doctoral Dissertation Fellowship.
The authors also wish to thank E. F. Polomski with initial assistance in
planning PID 131 activities, J. Crovisier for useful discussions,  
E. Peeters for providing her PAH spectra in digital form, and 
the referee who helped improve the discussion presented in the manuscript.

{\it Facilities:} \facility{Spitzer (IRS)}

\clearpage


\clearpage



\begin{deluxetable}{lcrcccc}
\tablewidth{0pt}
\tablecaption{WATER IN C/2003 K4 (LINEAR): MODEL FITS \label{table:tb_k4}}
\tablehead{
\colhead{Extraction}
& \colhead{Offset\tablenotemark{a}}
& \colhead{Offset\tablenotemark{a}}
& \colhead{Band Intensity\tablenotemark{b}}
& \colhead{$T_{rot}$\tablenotemark{c}} & \colhead{OPR\tablenotemark{c}}
& \colhead{$\chi^{2}_{\nu}$\tablenotemark{c,d}} \\
\colhead{Slit ID}
& \colhead{(\arcsec)}
& \colhead{(km)}
& \colhead{(10$^{-20}$ W cm$^{-2}$)}
& \colhead{($K$)} & 
}
\startdata
A 0-27 &   \phantom{0}7.2 &  7377 & 5.51 $\pm$ 0.16 & 33.5 $\pm$ 4.0
& 2.47 $ \pm$ 0.30 & 0.5 \\
B 0-22 &  12.9 & 13198 & 2.67 $\pm$ 0.10  &  28.6 $\pm$ 3.6 & 2.47 $\pm$ 0.34 &  1.0 \\
C 0-17 &  21.3 &  21785& 1.62 $\pm$ 0.09  &  22.8 $\pm$ 4.3 & 2.5\tablenotemark{e} & 1.6 \\
D 1-27 &  14.4 & 14682 & 2.88 $\pm$ 0.10  &  27.6 $\pm$ 3.5
& 2.13 $\pm$ 0.31 & 0.5 \\
E 1-22 & 18.0 & 18348   & 1.93 $\pm$ 0.09 & 20.9 $\pm$ 3.5  & 2.56 $\pm$ 0.47 & 1.5 \\
F 1-17 &  24.7 & 25229& 1.19 $\pm$ 0.09 &  24.8 $\pm$ 5.6  & 2.5\tablenotemark{e} & 1.6  \\
G 2-27 & 21.7 & 22154 & 1.68 $\pm$ 0.08 &  22.4 $\pm$ 4.2
& 2.61 $\pm$ 0.31 & 0.9 \\
H~2-22 & 24.2 & 24750 & 1.46 $\pm$ 0.08 & 23.2 $\pm$ 4.3   & 2.38 $\pm$ 0.48 & 1.5 \\
I~~2-17 & 29.6 & 30209 & 1.02 $\pm$ 0.09 &   17.7 $\pm$ 5.0  & 2.5\tablenotemark{e} & 2.4   \\
\enddata

\tablenotetext{a}{Relative to the comet C/2003 K4 (LINEAR) ephemeris
position derived from JPL \#96.}

\tablenotetext{b}{Band intensity above fitted continuum between
5.8 and 7.1~\micron.}

\tablenotetext{c}{Fits to the 6.3--7.0~\micron \ region.} 

\tablenotetext{d}{Reduced $\chi^2$ between 6.3 and 7.0~\micron \
($\chi^{2}_{\nu}$ = $\chi^2$/$dof$, with number of degrees of
freedom [$dof$] = 23--3 = 20). }

\tablenotetext{e}{Assumed value.}

\end{deluxetable}
\clearpage


\begin{deluxetable}{lccc}
\tablewidth{0pt}
\tablecaption{WATER IN C/2003 K4 (LINEAR): MODEL FITS USING SEGMENTS
\label{table:split-fits}}
\tablehead{
\colhead{Extraction} & \colhead{Regions Fit} 
& \colhead{$T_{rot}$} &\colhead{OPR} \\
\colhead{Slit ID} & \colhead{(\micron)} 
& \colhead{($K$)} &
}
\startdata
A 0-27 & 5.85  -- 7.00 & 30.7 $\pm$ 3.2 & 2.47 $\pm$ 0.27 \\
A 0-27 & 5.85  -- 6.30 & 36.5 $\pm$ 5.8 & 2.03 $\pm$ 0.40 \\
A 0-27 & 6.30  -- 7.00 & 33.5 $\pm$ 4.0 & 2.47 $\pm$ 0.30 \\
\cline{1-4} \\
B 0-22 & 5.85  -- 7.00 & 26.5 $\pm$ 3.3 & 2.95 $\pm$ 0.40 \\
B 0-22 & 5.85  -- 6.30 & 59.5 $\pm$ 8.8 & 2.83 $\pm$ 0.68 \\
B 0-22 & 6.30  -- 7.00 & 28.6 $\pm$ 3.6 & 2.47 $\pm$ 0.34\\
\cline{1-4} \\
D 1-27 & 5.85  -- 7.00 & 27.0 $\pm$ 2.9 & 2.07 $\pm$ 0.24 \\
D 1-27 & 5.85  -- 6.30 & 42.7 $\pm$ 7.6 & 1.55 $\pm$ 0.30 \\
D 1-27 & 6.30  -- 7.00 & 27.6 $\pm$ 3.5 & 2.13 $\pm$ 0.31\\
\cline{1-4} \\
G 2-27 & 5.85  -- 7.00 & 21.4 $\pm$ 3.2 & 2.67 $\pm$ 0.41 \\
G 2-27 & 5.85  -- 6.30 & 54\phantom{.0} $\pm$ 12\phantom{.} & 2.16 $\pm$ 0.57 \\
G 2-27 & 6.30  -- 7.00 & 22.4 $\pm$ 4.2 & 2.61 $\pm$ 0.31 \\
\enddata
\end{deluxetable}
\clearpage



\clearpage


\clearpage



\clearpage



\clearpage



\clearpage



\clearpage


\clearpage


\clearpage


\clearpage



\clearpage


\begin{thebibliography}{}

\bibitem[A'Hearn et al.(2005)]{ahearn05} A'Hearn, M.~F., et al.\ 2005, 
Science, 310, 258 

\bibitem[Barber et al.(2006)]{BT2} Barber, R.~J., Tennyson, J., Harris,
G.~J., \& Tolchenov, R.~N.\ 2006, \mnras, 368, 1087

\bibitem[Belton et al.(2006)]{belton06} Belton, M.~J.~S., et al.\ 2006, 
Icarus in press 

\bibitem[Bensch \& Bergin(2004)]{ben04} Bensch, F., \& Bergin, E.~A.,
2004, \apj, 615, 531

\bibitem[Bernard Salas et al.(2001)]{bernard01} Bernard Salas,
J., Pottasch, S.~R., Beintema, D.~A., \& Wesselius, P.~R.\ 2001, \aap,
367, 949

\bibitem[Biver(1997)]{biv97} Biver, N., 1997, PhD thesis, University
of Paris 7

\bibitem[Biver et al.(2007)]{biv07} Biver, N., et al.\ 2007, P\&SS, in press 

\bibitem[Biver et al.(1997)]{biver-hb97} Biver, N., et al.\ 2007, Earth, 
Moon, \& Planets, 78, 5

\bibitem[Bockel\'ee-Morvan(1987)]{dbm87} Bockel\'ee-Morvan, D.,
1987, \aap, 181, 169

\bibitem[Bockel\'ee-Morvan \& Crovisier(1989)]{domi89}
Bockel\'ee-Morvan, D., \& Crovisier, J.\ 1989, \aap, 216, 278

\bibitem[Bockel\'ee-Morvan et al.(2000)]{bock00} 
Bockel{\'e}e-Morvan, D., et al.\ 2000, \aap, 353, 1101 

\bibitem[Bockel{\'e}e-Morvan et al.(2004)]{boc04} 
Bockel{\'e}e-Morvan, D., Crovisier, J., Mumma, M.~J., \& Weaver, H.~A.\ 
2004, in Comets II, eds. M. Festou, H.~U. Keller, and H.~A. Weaver, 
(University of Arizona Press: Tucson), p.391


\bibitem[Bonev et al.(2007)]{bonev07} Bonev, B.~P., Mumma, M.~J.,
Villanueva, G.~L., Disanti, M.~A., Ellis, R.~S., Magee-Sauer, K.,
\& Dello Russo, N.\ 2007, \apj, 661, L97

\bibitem[Bonev et al.(2006)]{bonev06} Bonev, B.~P., Mumma, M.~J.,
Disanti, M.~A., Dello Russo, N., Magee-Sauer, K., Ellis, R.~S., 
\& Stark, D.~P.\ 2006, \apj, 653, 774

\bibitem[Capria(2002)]{capria02} Capria, M.~T.\ 2002, Earth, Moon, \& 
Planets, 89, 161 

\bibitem[Capria et al.(2000)]{capria00} Capria, M.~T., et al.\ 2000, 
\aap, 357, 359 

\bibitem[Chiar et al.(2000)]{chi00} Chiar, J.~E., Tielens, A.~G.~G.~M., 
Whittet, D.~C.~B., Schutte, W.~A., Boogert, A.~C.~A., Lutz, D., 
van Dishoeck, E.~F., \& Bernstein, M.~P.\ 2000, \apj, 537, 749


\bibitem[Crovisier(1989)]{cro89} Crovisier, J., 1989, \aap, 213, 459

\bibitem[Crovisier(2002)]{cro02} Crovisier, J.\ 2002, Constants for
molecules of astrophysical interest in the gas phase: photodissociation,
microwave and infrared spectra. http://www.lesia.obspm.fr/\~{}crovisier/basemole

\bibitem[Crovisier(2007)]{crovs07} Crovisier, J.\ 2007, arXiv:astro-ph/0703785  

\bibitem[Crovisier \& Bockel\'ee-Morvan(2007)]{cro07} Crovisier, J., 
\& Bockel\'ee-Morvan D. \ 2007, Icarus (in press)

\bibitem[Crovisier et al.(2000)]{crovs00} Crovisier, J., et al.\ 2002,
in Thermal Emission Spectroscopy and Analysis of Dust, ASP Conf. Ser. 196, 
eds. Y. Pendelton, D. Cruikshank [ASP: San Francisco], 109 

\bibitem[Crovisier et al.(1997a)]{cro97a} Crovisier, J., et al.\ 1997a,
First ISO Workshop on Analytical Spectroscopy, ESA SP 419, 137

\bibitem[Crovisier et al.(1997b)]{cro97b} Crovisier, J., et al.\ 1997b,
Science 275, 1904

\bibitem[Dello Russo et al.(2005)]{del05} Dello Russo, N., Bonev,
B.~P., DiSanti, M.~A., Mumma, M.~J., Gibb, E.~L., Magee-Sauer, K.,
Barber, R.~J., \& Tennyson, J.\ 2005, \apj, 621, 537


\bibitem[Dello Russo et al.(2004)]{del04} Dello Russo, N.,
DiSanti, M.~A., Magee-Sauer, K., Gibb, E.~L., Mumma, M.~J., 
Barber, R.~J., \& Tennyson, J.\ 2004, Icarus, 168, 186

\bibitem[Eberhardt \& Krankowsky(1995)]{eberhardt95} Eberhardt, P.,
\& Krankowsky, D.\ 1995, \aap, 295, 795

\bibitem[Ehrenfreund et al.(2004)]{ehren04} Ehrenfreund, P., et al.\ 
2004, in Comets II, eds. M. Festou, H.~U. Keller, and H.~A. Weaver, 
(University of Arizona Press: Tucson), p.115 

\bibitem[Gehrz et al.(2007)]{gehrz07} Gehrz, R.~D., Roellig, T.~L.,
Werner, M.~W., Fazio, G.~G., Houck, J.~R., Low, F.~J., Rieke, G.~H.,
Soifer, B.~T., Levine, D.~A., \& Romana, E.~A.\ 2007, Rev. Sci. Instr.,
78, 011302

\bibitem[Gehrz et al.(2005)]{gehrzet05}  Gehrz, R.~D., Hanner, M.~S.,
Homich, A.~A., \& Tokunaga, A.~T.\ 2005, \aj, 130, 2383

\bibitem[Gunnarsson et al.(2003)]{gunna03} Gunnarsson, D., et al.\ 
2003, \aap, 402, 383 

\bibitem[Harker et al.(2007)]{harker07} Harker, D.~E., Woodward, C.~E., 
Wooden, D.~H., Trujillio, C., \& Fisher, S.\ 2007, Icarus, in press 

\bibitem[Jacquinet-Husson et al.(2005)]{jacq05}
Jacquinet-Husson, N., et al.\ 2005, JQRST, 95, 429

\bibitem[Houck et al.(2004)]{houck04} Houck, J.~R., et al.\ 2004,
\apjs, 154, 18

\bibitem[Kawakita et al.(2006)]{kawakita06} Kawakita, H., et al.\ 2006,
\apj, 643, 1337

\bibitem[Keller et al.(2006)]{kel06} Keller, L.~P., et al.\ 2006, 
Science, 314, 1728

\bibitem[Kelley et al.(2006)]{kelley06} Kelley, M.~S., et al.\ 2006,
\apj, 651, 1256


\bibitem[Lisse et al.(2007)]{lisse07} Lisse, C.~M., Kraemer, K.~E., 
Nuth, J.~A., Li, A., \& Joswiak, D.\ 2007, Icarus 187, 69

\bibitem[Lisse et al.(2006)]{lisse06} Lisse, C.~M., et al.\ 2006,
Science, 313, 635

\bibitem[Markwick \& Charnley(2005)]{marcharn04} Markwick, A.~J. 
\& Charnley, S.~B.\ 2005, in Highlights of Astronomy, Vol.~13, 
eds. O. Engvold, (Astron. Soc. Pacific: San Francisco) p.518

\bibitem[Meech \& Sovern(2004)]{meechs04} Meech, K. \& Svoren, J.\ 
2004, in Comets II, eds. M. Festou, H.~U. Keller, and H.~A. Weaver, 
(University of Arizona Press: Tucson), p.317 

\bibitem[Mumma et al.(2003)]{mum03} Mumma, M. J., et al. 2003, Adv. 
Space Res., 31, 2563 

\bibitem[Mumma et al.(1993)]{mum93} Mumma, M.~J., Weissman, P.~R., \&
Stern, S.~A.\ 1993, Protostars and Planets III, eds. M.~H. Levy
and J.~I. Lunine, (University of Arizona Press: Tucson), p.1177


\bibitem[Or\'o et al.(2006)]{oro06} Or\'o, J., Lazcano, A., \& 
Ehrenfreund, P.\ 2006, in Comets and the Origin and Evolution of Life, 
eds. P.~J. Thomas, R.~D. Hicks, C.~F. Chyba, C.~P. McKay (Springer: New 
York), p.1 

\bibitem[Partridge \& Schwenke(1997)]{PS97} Partridge, H., \& Schwenke,
D.~W.\ 1997, J. Chem. Phys., 109, 4618

\bibitem[Peeters et al.(2002)]{pee02} Peeters, E., Hony, S., 
Van Kerckhoven, C., Tielens, A.~G.~G.~M., Allamandola, L.~J., 
Hudgins, D.~M., \& Bauschlicher, C.~W.\ 2002, \aap, 390, 1089 

\bibitem[Pendleton \& Allamandola(2002)]{pen02} Pendleton, Y.~J., 
\& Allamandola, L.~J.\ 2002, \apjs, 138, 75 


\bibitem[Prialnik et al.(2004)]{prialnik04} Prialnik, D., Benkhoff, J., 
\& Podolak, M.\ 2004, in Comets II, eds. M. Festou, H.~U. Keller, and 
H.~A. Weaver, (University of Arizona Press: Tucson), p.359 

\bibitem[Prialnik(2002)]{prialnik02} Prialnik, D.\ 2002, Earth, Moon, \& 
Planets, 89, 27 

\bibitem[Rothman et al.(2004)]{roth05}
Rothman, L.S., et al.\ 2005, JQSRT, 96, 139

\bibitem[Russell et al.(2004)]{iauc8361}  Russell, R.~W., Lim, D.~L., 
Sitko, M.~L., \& Carpenter, W.~J.\ 2004, IAUC, 8361

\bibitem[Schulz, St\"{u}we, \& Erd(2005)]{shulz05} Schulz, R., St\"{u}we, 
J.~A., \& Erd, C.\ 2005, Earth, Moon, \& Planets, 97, 387

\bibitem[Sitko et al.(2004)]{iauc8391} Sitko, M.~L., Russell, R.~W., 
Lynch, D.~K., \& Lim, D.~L.\ 2004, IAUC, 8391

\bibitem[Smith et al.(2007)]{smith07} Smith, J.~D. et al.\  2007, 
\pasp, submitted

\bibitem[Spitzer Science Center(2006)]{irsdh} \spitzer{} Science
Center\ 2006, Infrared Spectrograph Data Handbook (Pasadena: SSC)
http://ssc.spitzer.caltech.edu/irs/dh/dh30\_v1.pdf/

\bibitem[Spoon et al.(2000)]{spoo00} Spoon, H.~W.~W., Koornneef, J., 
Moorwood, A.~F.~M., Lutz, D., \& Tielens, A.~G.~G.~M.\ 2000, \aap, 357, 898 

\bibitem[Werner et al.(2004)]{werner04} Werner, M.~W., et al.\
2004, \apjs, 154, 1

\bibitem[Wooden et al.(2004)]{wooden04} Wooden, D.~H., et al.\ 2005, in 
Comets II, eds. M. Festou, H.~U. Keller, and H.~A. Weaver, (University 
of Arizona Press: Tucson), p.33

\bibitem[Woodward et al.(2007)]{woodward07} Woodward, C.E., Kelley, M.~S., 
Bockel{\'e}e-Morvan, D., et al.\ 2007, in preparation

\bibitem[Woodward et al.(2004)]{iauc8378} Woodward, C.~E., Kelley, M.~S., 
\& Wooden, D.~H.\ 2004, IAUC, 8131

\bibitem[Xie \& Mumma(1992)]{xie92} Xie, X., \& Mumma, M.J., 1992,
\apj, 386,720

\bibitem[Young \& McGaha(2004)]{iauc8131} Young, J. \& McGaha, J.\ 2004, 
IAUC, 8131

\bibitem[Zakharov et al.(2007)]{zak07} Zakharov,V., et al., 2007,
\aap, in press
\end{thebibliography}
\end{document}